\documentclass[11pt,a4paper]{article}

\usepackage{graphicx}
\usepackage{dcolumn}
\usepackage{bm}
\usepackage{caption}
\usepackage{epsfig,multicol,bbm}
\usepackage{amsmath,amssymb}
\usepackage[labelsep=colon]{caption}

\newcommand\fverb{\setbox\fverbbox=\hbox\bgroup\verb}
\newcommand\fverbdo{\egroup\medskip\noindent%
            \fbox{\unhbox\fverbbox}\ }
\newcommand\fverbit{\egroup\item[\fbox{\unhbox\fverbbox}]}
\newbox\fverbbox

\begin{document}

\title{{Investigating the effect of nuclear deformation on $\beta$-decay half-lives of neutron-rich nuclei}
\author{Jameel-Un Nabi, Tuncay Bayram, Wajeeha Khalid, Arslan Mehmood*, \\ Alper K\"oseo\u{g}lu}\\}
 \maketitle

\begin{abstract}
We examine the effect of nuclear deformation on the calculated  $\beta$-decay half-lives of 55 neutron-rich nuclei. The deformation values were computed using DD-PC1 and DD-ME2 interactions in the Relativistic Hartree-Bogoliubov model. Yet another set of deformation was adopted from the Finite Range Droplet Model (FRDM). The model-dependent deformation values were used as a free parameter in the proton-neutron quasiparticle random phase approximation  model to calculate the $\beta$-decay properties under terrestrial and stellar conditions. The Gamow-Teller  strength distributions, branching ratios, half-lives and stellar weak rates of selected nuclei  were later investigated. It was concluded that the $\beta$-decay properties of neutron-rich nuclei changed with nuclear deformation.  The FRDM computed deformation values provided the best predictions for calculated half-lives followed by the DD-ME2 functional. The terrestrial  $\beta$-decay half-lives changed up to 3 orders of magnitude and the stellar $\beta$-rates within a factor of 2 as the neutron-rich nuclei switched their geometrical configurations.   For magic number nucleus $^{138}$Sn, the stellar rates changed substantially by more than 1 order of magnitude as the nucleus transitioned away from the spherical shape. Our investigation might prove useful for a realistic modeling of nucleosynthesis calculation. 
\end{abstract}

\section{Introduction}
The evolutionary phases of stars and the nucleosynthesis processes are intricately connected. The elements created in stars get recycled into the space to be later transformed into new planets and stars during the universe's roughly 13.8 billion years old. Approximately half of the heavy elements in nature are created by the rapid neutron capture process ($r$-process), occurring near the neutron-drip line. Due to limited experimental data, theoretical predictions are essential for understanding these nuclear properties. While traditionally linked to supernovae, recent studies suggest neutron-rich conditions might also arise from neutron star matter ejections or rapid ejection of matter undergoing intense electron captures \cite{Cow21}. The $r$-process~\cite{Bur57}  yields a big proportion of all massive nuclei in the universe. The magic neutron number nuclei with $N= 126, ~82$ and $50$ play an important part for the neutron capture process \cite{Cow91}. Half-lives of magic neutron number nuclei $N$ = 50 and 82 were measured earlier \cite{Pfe01}. In recent years, examination of the nuclear structure properties of neutron-rich nuclei, in the mass range 110 $\leq A \leq$ 120 and heavier~\cite{Woo92, Hey11}, gained attention of researchers both theoretically and experimentally.  This region is characterized by rapid modifications in structure of the ground as well as low-lying transition states. Both relativistic \cite{Xia12, Mei12} and non-relativistic \cite{Ben09, Rod11} analyses of the nuclear structural evolution  reveal that the equilibrium shape of the nucleus rapidly varies with the number of nucleons in this mass region. At near energies, there is shape coexistence with competing triaxial, axially deformed (oblate and prolate) and spherical shapes.

The $\beta$-decay properties e.g., branching ratios, Gamow-Teller (GT) strengths, half-lives are important in the study of simulation of astrophysical events including core-collapse supernovae. The GT strength functions and $\beta$-decay half-lives were calculated using various models including large scale shell model (e.g.,~\cite{Lan00}), continuum quasiparticle random phase approximation (e.g.,~\cite{Bor96}), gross theory (e.g.,~\cite{Tak69}) and proton-neutron quasiparticle random phase approximation  (e.g.,~\cite{Mut89}). The $\beta$-decay properties of neutron-rich nuclei $^{77-79}$Cu, $^{79, 81}$Zn and $^{82}$Ga were measured at the National Superconducting Cyclotron Laboratory using the $\beta$-counting station in conjunction with the neutron detector NERO~\cite{Hos10}. More recently, the $\beta$-decay properties were calculated for neutron-rich Cerium isotopes in the mass region $120 \leq A \leq 157$ within the pn-QRPA approach \cite{Nab22}. The $\beta$-decay half-lives of $^{97-103}$Rb, $^{98-107}$Sr, $^{101-109}$Y, $^{104-112}$Zr were computed using the same model \cite{She22}. The thermal QRPA was used to examine the $\beta$-decay rates with thermal effect in the past \cite{Dzh08}. It was noted that $\beta$-decay rates increase as the GT fragmentation got redistributed with temperature effect. Soaring temperatures led to increase in the fraction of the GT strength lying below the thermal vacuum. The thermal effect  on $\beta$-decay half-lives for even-even isotones with neutron magic number $N$ = 82 was investigated using  the finite-temperature quasiparticle random phase approximation  on the basis of the finite temperature Skyrme-Hartree-Fock + Bardeen-Cooper-Schrieffer (BCS) method~\cite{Min09}. It was concluded that for all nuclei, the $\beta$-decay half-lives decreased gradually as the temperature increased. It was suggested that the thermal effect on calculated half-lives was negligible at the  standard $r$-process temperature ($<$ 0.2 MeV). A microscopic nuclear model that can precisely predict the charge-changing matrix elements in accordance with the available measured data is required  to examine the GT transitions of unstable nuclei. 

One such effective model, known as the proton-neutron quasiparticle random phase approximation (pn-QRPA) \cite{Hal16} is widely used for a reliable calculation of $\beta$-decay properties of unstable nuclei. 
Approximately 6000 nuclei exist between neutron-drip and  stability line. The $\beta$ transitions occur at high density, $\rho \geq 10^3$ gcm$^{-3}$, and temperature T $ \geq 10^3$ K \cite{Tak89} during the $r$- and $s$-processes (slow neutron-capture processes). Using microscopic nuclear theory, Klapdor was the first to calculate the $\beta$-decay rates of all nuclei beyond the stability line \cite{Kla84}. The calculations were later refined, using the pn-QRPA model, both for terrestrial \cite{Sta90, Hir93, Hom96} and  stellar environments \cite{Nab99, Nab04}. The model reproduced 93$\%$ of measured $\beta^+$ half-lives shorter than 60 $s$ within a factor of 10~\cite{Hir93}. It was shown that the accuracy of the nuclear model increased with increasing neutron excess and  96$\%$ of measured $\beta^-$ half-lives shorter than 60 $s$ were successfully reproduced within a factor of 10~\cite{Sta90}. The nuclear shell model often provides  accurate predictions for $\beta$-decay half-lives, especially for nuclei near closed shells. This model takes into account detailed nuclear structure effects by explicitly considering individual nucleon interactions within a well-defined potential. The pn-QRPA is capable of handling a wide range of nuclei, including those far from stability, because it treats collective excitation and correlations between protons and neutrons more efficiently. The pn-QRPA provides less detailed information about individual nuclear states and is more focused on collective excitation.

The $\beta$-decay half-lives for neutron-rich $^{134-139}$Sn, $^{148-153}$Ba, $^{137-144}$Te, $^{145-151}$Cs, $^{134-142}$Sb $^{140-146}$I,  $^{151-155}$La and $^{142-148}$Xe nuclei were recently calculated at the Radioactive Isotopes Beam factory (RIBF), RIKEN ~\cite{Wu20}. The investigation included 13 nuclei for which no prior experimental data was available.  Motivated from this work, we present here the $\beta$-decay properties of these neutron-rich nuclei using the theoretical framework of deformed pn-QRPA model. Our study consists of investigation of GT strength distributions, $\beta$-decay terrestrial half-lives,  and stellar rates of neutron-rich nuclei having mass $134 \leq A \leq 155$. The covariant density functional theory (CDFT)~\cite{rei89,rin96,bay13,typ18}, with density dependent interactions (DD-PC1 and DD-ME2), was used to calculate the nuclear deformation ($\beta_2$) values which is an important input parameter for the pn-QRPA calculation. In our investigation, yet another set of deformation was adopted from the Finite Range Droplet Model (FRDM) \cite{Mol16}. These three sets of deformation values  were used as input parameter in the pn-QRPA model  to examine the GT strength distributions, branching ratios, partial and total half-lives (terrestrial and stellar). 

We present the formalism of nuclear models used in Section 2. The examination of results is shown in Section 3. Summary and conclusion of current investigation are given in the last section.

\section{Theoretical Formalism}
The nuclear structure calculation was performed using the meson-exchange and point coupling models of the CDFT in Relativistic Hartree-Bogoliubov (RHB) formalism. The $\beta$-decay properties were calculated using the pn-QRPA model. Below we describe the essential formalism of the two models.
\subsection{Covariant Density Functional Theory (CDFT)}

\label{CDFT}

The first relativistic approach namely relativistic mean field theory (RMFT) for describing the structure and behavior of atomic nuclei was initially introduced within the context of quantum field theory~\cite{wal74}. It is an extension of the mean field theory that takes into account special relativity, unlike the non-relativistic mean field models like the Skyrme model. The model assumes that  nucleons move in atomic nucleus in an effective potential generated by the interactions with other nucleons. The key aspect is the inclusion of relativistic effects, accounting for high velocities and energies of nucleons within the nucleus~\cite{rin96}. After addition of density dependence, because of necessities for an accurate description of the properties of nuclear surface~\cite{bog77}, it was started to called as CDFT. Mainly, interactions of nucleons in the CDFT are based on exchanging of mesons between nucleons in atomic nuclei. In this theory, the saturation mechanism are produced via $\sigma$- and $\omega$-meson fields. The scalar $\sigma$-meson and vector $\omega$-meson fields are related with attractive and repulsive part of nuclear interaction, respectively. The isovector $\rho$-meson provides asymmetry component. Three versions of CDFT are mainly used for calculations based on the consideration of meson-nucleon interaction and self-interaction of mesons (see Ref.~\cite{men06} and references therein). In the present study, two types of CDFT, meson-exchange and point-coupling, have been taken into account within the framework of the RHB formalism for the calculation of electric quadrupole deformation parameter ($\beta_2$) of the considered nuclei. In this part, the meson-exchange and point coupling version of the CDFT are described.  
\subsubsection{Meson-exchange Model}
In this model, a phenomenological Lagrangian density consists of three parts which can be written as

\begin{equation}
	\mathcal{L}=\mathcal{L}_N+\mathcal{L}_m+\mathcal{L}_{int},\label{lagden}
\end{equation}  
where $\mathcal{L}_N$ is a term for the free nucleon while $\mathcal{L}_m$ term covers meson and electromagnetic fields. The term $\mathcal{L}_{int}$ includes meson-nucleon and photon-nucleon interactions. Their explicit forms are gives as 

\begin{equation}
	\mathcal{L}_N=\bar{\Psi}(i\gamma_\mu\partial^\mu-m)\Psi,\label{lagnuc}
\end{equation}
where $\psi$ and $m$ indicates the Dirac spinor and the nucleon mass, respectively.
\begin{eqnarray}
	\mathcal{L}_m &=\frac{1}{2}\partial_\mu\sigma\partial^\mu\sigma-\frac{1}{2}m_\sigma^2\sigma^2-\frac{1}{2}\Omega_{\mu\nu}\Omega^{\mu\nu}+\frac{1}{2}m^2_\omega\omega_\mu\omega^\mu \nonumber \\ 
	&-\frac{1}{4}\overrightarrow{R}_{\mu\nu}.\overrightarrow{R}^{\mu\nu}+\frac{1}{2}m_\rho^2 \overrightarrow{\rho}_\mu.\overrightarrow{\rho}^\mu -\frac{1}{4}F_{\mu\nu}F^{\mu\nu}, \label{lagmes}
\end{eqnarray} 
where arrows indicate isovectors. $m_\sigma$, $m_\omega$ and $m_\rho$, represent the masses of the related mesons while $\Omega_{\mu\nu}$, $\overrightarrow{R}_{\mu\nu}$ and $F_{\mu\nu}$ are field tensors. 
\begin{eqnarray}
	\mathcal{L}_{int}&= -g_\sigma\bar{\Psi}\Psi \sigma - g_\omega\bar{\Psi}\gamma^\mu\Psi\omega_\mu - g_\rho\bar{\Psi}\overrightarrow{\tau}\gamma^\mu\Psi.\overrightarrow{\rho}_\mu \nonumber \\ 
	&-e\bar{\Psi}\gamma^\mu\Psi A_\mu,
	\label{lagrangian}
\end{eqnarray}
where the coupling constants of the related mesons are indicated by $g_\sigma$, $g_\omega$ and $g_\rho$. 

Usage of the Lagrangian density yields the Hamiltonian density for the static case~\cite{rin96},
\begin{eqnarray}
	\mathcal{H}({\bf r}) &=\Sigma_i^\dagger(\bm{\alpha p}+\beta m)\Psi_i \nonumber\\
	&+ \frac{1}{2}\left[ ({\bf \nabla}\sigma)^2 + m_\sigma^2\sigma^2\right] - \frac{1}{2}\left[ ({\bf \nabla}\omega)^2 + m_\omega^2\omega^2\right]\nonumber \\ 
	&-\frac{1}{2}\left[ ({\bf \nabla}\rho)^2 + m_\rho^2\rho^2\right] - \frac{1}{2}({\nabla \bf A})^2 \nonumber\\
	&+ \left[ g_\sigma\rho_s\sigma+g_\omega j_\mu\omega^\mu + g_\rho \overrightarrow{j}_{\mu}.\overrightarrow{\rho}^{\mu} + ej_{p\mu}A^{\mu}  \right], \label{hamden}
\end{eqnarray} 
where $\rho_s({\bf r})$, $j_\mu({\bf r})$, $\overrightarrow{j}_\mu({\bf r})$ and $j_{p\mu}({\bf r})$ indicates the isoscalar-scalar density, isoscalar-vector, isovector-vector and electromagnetic currents, respectively. Their explicit forms are given as  
\begin{equation}
	\rho_s({\bf r})=\Sigma_{i=1}^A \bar{\Psi}_i({\bf r}) \Psi_i({\bf r}), \label{isoscalar}
\end{equation}

\begin{equation}
	j_\mu({\bf r})=\Sigma_{i=1}^A \bar{\Psi}_i({\bf r}) \gamma_\mu \Psi_i({\bf r}), \label{isovector}
\end{equation}

\begin{equation}
	\overrightarrow{j}_\mu({\bf r})=\Sigma_{i=1}^A \bar{\Psi}_i({\bf r}) \overrightarrow{\tau}\gamma_\mu \Psi_i({\bf r}), \label{isovectorvec}
\end{equation}

\begin{equation}
	j_{p\mu}({\bf r})=\Sigma_{i=1}^Z \bar{\Psi}_i^\dagger({\bf r})\gamma_\mu \Psi_i({\bf r}). \label{electromagnetic}
\end{equation}

The {\it no sea approximation}  can be used for summation in these densities and the integration of the Hamiltonian density produces the total energy as follows
\begin{equation}
	E\left[\Psi,\bar{\Psi},\sigma,\omega^{\mu},\overrightarrow{\rho}^{\mu}, A^{\mu} \right]=\int d^3r \mathcal{H}({\bf r}). \label{toten}
\end{equation}     
The variation of energy density functional given in Eq.~(\ref{toten}) leads to the Dirac equation,
\begin{equation}
	\widehat{h}_D \Psi_i=\epsilon_i\Psi_i, \label{diraceq}
\end{equation} 
and a set of differential equations given by 
\begin{equation}
	\left[ -\Delta + m_\sigma^2 \right]\sigma=-g_\sigma\rho_s , \label{helm1}
\end{equation} 
\begin{equation}
	\left[ -\Delta + m_\omega^2 \right]\omega^\mu=g_\omega j^\mu , \label{helm2}
\end{equation}
\begin{equation}
	\left[ -\Delta + m_\rho^2 \right]\overrightarrow{\rho}^\mu=g_\rho \overrightarrow{j}^m , \label{helm3}
\end{equation}
\begin{equation}
	-\Delta A^\mu = ej_p^\mu. \label{helm4}
\end{equation} 

In Eq.~(\ref{diraceq}),  $\widehat{h}_D$ is the Dirac Hamiltonian given by
\begin{equation}
	\widehat{h}_D=\bm{\alpha}(\bm{p-\Sigma}) + \Sigma_0 + \beta(m+\Sigma_s). \label{dirham}
\end{equation} 
The explicit forms of the single-nucleon self-energies represented by the term $\Sigma$ in Eq. (\ref{dirham}) are given as
\begin{eqnarray}
	\Sigma_s({\bf r}) &=g_\sigma\sigma ({\bf r}), \\ 
	\Sigma_\mu ({\bf r}) &=g_\omega\omega_\mu ({\bf r}) + g_\rho\overrightarrow{\tau}.\overrightarrow{\rho}_\mu({\bf r}) + eA_\mu({\bf r})\nonumber\\ &+ \Sigma_\mu^R({\bf r}). \label{selfener}
\end{eqnarray}
The rearrangement contribution to the vector self-energy can be written as

\begin{equation}
	\Sigma_\mu^R=\frac{j_\mu}{\rho_v}\left( \frac{\partial g_\sigma}{\partial \rho_v}\rho_s\sigma + \frac{\partial g_\omega}{\partial \rho_v} j_v\omega^v + \frac{\partial g_\rho}{\partial \rho_v}\overrightarrow{j}_v.\overrightarrow{\rho}^v \right). \label{rear}
\end{equation}    

For the solution of even-even atomic nuclei, there are no currents. The spatial components of the meson field could be vanished. Thus, the Dirac equation given in Eq.~(\ref{diraceq}) transforms to the following equation

\begin{equation}
	\left\lbrace -i{\bm \alpha\nabla}+\beta M^*({\bf r}) + V({\bf r})\right\rbrace \Psi_i({\bf r})=\epsilon_i \Psi_i({\bf r}) , \label{mass1}
\end{equation}
where $M^*({\bf r})$ represents the effective mass given by  

\begin{equation}
	M^*({\bf r})=m+g_\sigma \sigma,
\end{equation}
and $V({\bf r})$ denotes the vector potential given by
\begin{equation}
	V({\bf r})= g_\omega \omega + g_\rho \tau_3 \rho +eA_0 + \Sigma_0^R. \label{vecpot}
\end{equation}
The rearrangement contribution in Eq.~(\ref{rear}) can be reduced as follows 
\begin{equation}
	\Sigma_0^R=\frac{\partial g_\sigma}{\partial \rho_v}\rho_s\sigma + \frac{\partial g_\omega}{\partial \rho_v} \rho_v\omega + \frac{\partial g_\rho}{\partial \rho_v}\rho_{tv}\rho, \label{rear2}
\end{equation} 
where the isovector density is denoted by $\rho_{tv}$. The details of parametrization for the meson-nucleon couplings could be found in Ref.~\cite{nik02b,typ05}.


\subsubsection{Point-coupling Model}

\label{PcM}

In the density-dependent point-coupling model, the isoscalar-scalar $\sigma$ meson, the isoscalar-vector $\omega$ meson, and the isovector-vector $\rho$ meson are taken into account~\cite{nik08}. The effective Lagrangian density which includes the free-nucleon term, the point-coupling interaction terms, and proton-electromagnetic field coupling term is represented by  

\begin{equation}
	\begin{split}
		\mathcal{L}&=\bar{\Psi}(i\gamma.\partial-m)\Psi \\
		&-\frac{1}{2}\alpha_S(\rho)(\bar{\Psi}\Psi)(\bar{\Psi}\Psi) -\frac{1}{2}\alpha_v(\rho)(\bar{\Psi}\gamma^\mu\Psi)(\bar{\Psi}\gamma_\mu\Psi)\\ 
		&-\frac{1}{2}\alpha_{TV}(\rho)(\bar{\Psi}\overrightarrow{\tau}\gamma^\mu\Psi)(\bar{\Psi}\overrightarrow{\tau}\gamma_\mu\Psi)\\
		&-\frac{1}{2}\delta_S(\partial_v\bar{\Psi}\Psi)(\partial^v\bar{\Psi}\mu\Psi)-e\bar{\psi}\gamma.A\frac{1-\tau_3}{2}\Psi. \label{lagdenp}
	\end{split}
\end{equation}

The derivative terms in Eq.~(\ref{lagdenp}) account for the leading effects of finite-range interactions. Conventionally, the functional form of couplings,  

\begin{equation}
	\alpha_i=a_i+(b_i+c_ix)e^{-d_ix},~~~(i=S, V, TV) \label{pcoup}
\end{equation}
is used. In Eq.~(\ref{pcoup}), $\rho_{sat}$ represent the nucleon density at saturation in symmetric nuclear matter and $x=\rho/\rho_{sat}$. For details of density-dependent point-coupling model we refer to ~\cite{nik08}.

\subsubsection{Details of RHB Calculations}

\label{RHB}

In the present study, we have performed axially deformed RHB calculations~\cite{nik14} for the considered nuclei using both density-dependent meson-exchange and point-coupling models of the CDFT. The RHB equations were solved in configurational space of harmonic oscillator wave functions and densities in coordinate space. In this study, 14 and 20 harmonic oscillator shells for fermionic and bosonic expansions were used. The used code DIRHB~\cite{nik14} for calculation of deformation parameter ($\beta_2$) of the considered nuclei starts with guessing values of the fields to determine potential terms which can will be used the solution of Dirac equation later. The solution of Dirac equation is used to calculate the sources which will be used to solution of Klein-Gordon type equations. These solutions are used for the next iteration which starts with the solution of the Dirac equation. The iterative procedure is cut after the convergence. 

The $\beta_2$ values for the considered nuclei in the RHB calculations is obtained from the calculated quadrupole moments given by 
\begin{equation}
	\beta_2=\frac{Q}{AR_0^2}\sqrt{\frac{5\pi}{9}}, \label{quad3}
\end{equation}
where $R_0=1.2A^{1/3}$ (fm) and $Q$ is the sum of quadrupole moments for neutrons and protons ($Q=Q_n+Q_p$). $Q_n$ and $Q_p$ are calculated by 
\begin{equation}
	Q_{n,p}=\langle 2r^2 P_2(cos\theta)\rangle_{n,p}=\langle 2z^2-x^2-y^2\rangle_{n,p}, \label{quad1}
\end{equation}
which are related to the expectation values of the spherical harmonics given by the following equation 
\begin{equation}
	\langle r^2 Y_{20}(\theta,0)\rangle_{n,p}=\frac{1}{2}\sqrt{\frac{5}{4\pi}}Q_{n,p}. \label{quad2}
\end{equation}

Pairing correlations are important for all open-shell nuclei. In our calculations, separable pairing force was used for treatment of pairing correlations. The considered density dependent models are DD-PC1 (point-coupling)~\cite{nik08} and DD-ME2 (meson-exchange)~\cite{lal05}.

\subsection{The pn-QRPA model}

\label{pn-QRPA}

This model was used to calculate the GT strength functions and $\beta$-decay half-lives for the neutron-rich nuclei. Following form of the Hamiltonian was selected for solution
\begin{equation} \label{eq24}
	~  ~ ~ ~ ~ ~ ~ ~ ~ ~ ~	\mathcal{H}^{QPRA} = \mathcal{H}^{sp} + \mathcal{V}^{pair} + \mathcal{V}^{pp}_{GT} + \mathcal{V}^{ph}_{GT} ,
\end{equation}
where the first term on the right side of Eq.~\ref{eq24} is the single particle Hamiltonian. $\mathcal{V}^{pair}$ symbolizes the pairing force. The remaining two terms  $\mathcal{V}_{GT}^{pp}$($\mathcal{V}_{GT}^{ph}$) show the particle-particle (particle-hole) GT
force.  The Nilsson model was used  to evaluate the values of single particle wavefunctions and energies~\cite{Nil55}. The oscillator constant  was calculated using the  expression,
$\hbar\omega=\left(\frac{45}{A^{1/3}}-\frac{25}{A^{2/3}}\right)$MeV (same for neutrons and protons). We used Nilsson-potential parameters from Ref. \cite{Rag84}.  The experimental $Q$-values were adopted from Ref.~\cite{Kon21}.  The nuclear deformation is one of the important parameters in our chosen model. Three different values of $\beta_2$ were used in our investigation. One of the selected $\beta_2$ values was adopted from the FRDM model \cite{Mol16}. The remaining two were calculated using interactions (DD-PC1 and DD-ME2) of RHB formalism (see previous subsection).

The pairing force was calculated  employing the BCS approximation. The BCS calculations were performed separately for both the neutron and proton systems. We assumed G as a constant 
pairing force ($G_p$ and $G_n$, for protons and neutrons, respectively). The pairing force was calculated using
\begin{eqnarray}\label{eq25}
	\mathcal{V}_{pair}=-G\sum_{jkj^{'}k^{'}}(-1)^{l+j-k}s^{\dagger}_{jk}s^{\dagger}_{j-k}
	\nonumber\\~~~~~~~~~~~~~~~~~~~~~~\times(-1)^{l^{'}+j^{'}-k^{'}} s_{j^{'}-k^{'}}s_{j^{'}k^{'}},
\end{eqnarray}
where  $l$ is orbital angular momentum and summation over $k$ and $k{'}$ was restricted to positive values. The spherical nucleon basis,  represented by ($s^{\dagger}_{jk}$, $s_{jk}$),  was transformed into deformed basis ($d^{\dagger}_{k\alpha}$, $d_{k\alpha}$) using the equation
\begin{equation}\label{eq26}
	~ ~ ~ ~ ~ ~ ~ ~ ~ ~ ~ ~ ~ ~ ~ ~ ~	d^{\dagger}_{k\alpha}=\Sigma_{j}D^{k\alpha}_{j}s^{\dagger}_{jk}.
\end{equation}
The transformation matrix $D$ is a set of Nilsson eigenfunctions and $\alpha$ represents the additional quantum numbers.  Later, we used the Bogoliubov transformation to introduced the quasiparticle basis $(q^{\dagger}_{k\alpha}, q_{k\alpha})$ 
\begin{equation}\label{eq27}
	~ ~ ~ ~ ~ ~ ~ ~ ~ ~ ~ ~	q^{\dagger}_{k\alpha}=u_{k\alpha}d^{\dagger}_{k\alpha}-v_{k\alpha}d_{\bar{k}\alpha}
\end{equation}
\begin{equation}\label{eq28}
	~ ~ ~ ~ ~ ~ ~ ~ ~ ~ ~ ~	q^{\dagger}_{\bar{k}\alpha}=u_{k\alpha}d^{\dagger}_{\bar{k}\alpha}+v_{k\alpha}d_{k\alpha},     (k > 0)    
\end{equation}
where the time-reversed state of $k$ was represented by $\bar{k}$. The occupation amplitudes satisfied the condition $v^{2}_{k\alpha}$ + $u^{2}_{k\alpha}$ = 1, and were computed using the BCS equations.

In our model, the GT transitions were expressed as creation of phonons. The pn-QRPA phonons were described using 
\begin{equation}\label{eq29}
	~ ~ ~ ~ ~ ~	{A}^{\dagger}_{\omega}(\mu)=\sum_{pn}[X^{pn}_{\omega}(\mu)q^{\dagger}_{p}q^{\dagger}_{\overline{n}}-Y^{pn}_{\omega}(\mu)q_{n}q_{\overline{p}}],
\end{equation}
where  $n$ and $p$ indices represent the neutron and proton quasiparticle states, respectively. The summation was over proton-neutron pairs satisfying the conditions: $\mu=k_{p}-k_{n}$ = (-1, 0, 1) and $\pi_{p}.\pi_{n}$=1. The excitation energy $\omega$ is the eigenvalue of the  RPA equation and (X$_{\omega}$, Y$_{\omega}$) are the amplitudes  of the pn-QRPA phonon.

In the pn-QRPA system, the proton-neutron residual interactions take place through the \textit{pp} and \textit{ph} GT forces, which were characterized by  interaction constants  $\kappa$ and $\chi$, respectively. In order to calculate the $ph$ GT force, we used the equation
\begin{equation}\label{eq30}
	~ ~ ~ ~ ~ ~ ~ ~ ~ ~ ~ ~	\mathcal{V}^{ph}= +2\chi\sum^{1}_{\mu= -1}(-1)^{\mu}\mathcal{U}_{\mu}\mathcal{U}^{\dagger}_{-\mu},
\end{equation}
with
\begin{equation}\label{eq31}
	\mathcal{U}_{\mu}= \sum_{j_{p}k_{p}j_{n}k_{n}}<j_{p}k_{p}\mid
	t_- ~\sigma_{\mu}\mid
	j_{n}k_{n}>s^{\dagger}_{j_{p}k_{p}}s_{j_{n}k_{n}}.
\end{equation}
The $pp$ GT force was explained using the expression
\begin{equation}\label{eq32}
	~ ~ ~ ~ ~ ~ ~ ~ ~ ~ ~ ~	\mathcal{V}^{pp}= -2\kappa\sum^{1}_{\mu=-1}(-1)^{\mu}\mathcal{O}^{\dagger}_{\mu}\mathcal{O}_{-\mu},
\end{equation}
with
\begin{eqnarray}\label{eq33}
	\mathcal{O}^{\dagger}_{\mu}= \sum_{j_{p}k_{p}j_{n}k_{n}}<j_{n}k_{n}\mid
	(t_- \sigma_{\mu})^{\dagger}\mid
	j_{p}k_{p}>\nonumber\\~~~~~~~~~~~~~~~~~~~~~~\times (-1)^{l_{n}+j_{n}-k_{n}}s^{\dagger}_{j_{p}k_{p}}s^{\dagger}_{j_{n}-k_{n}}.
\end{eqnarray}
The $\kappa$ and $\chi$ interaction strengths were determined using the relation $0.58/A^{0.7}$ and $5.2/A^{0.7}$, respectively, taken from Ref.~\cite{Hom96}. Our results fulfilled the model-independent Ikeda sum rule \cite{Ike63}. The reduced GT transition probabilities were calculated using
\begin{equation}\label{eq34}
	~ ~ ~ ~ ~ ~ ~ ~ 	\mathcal{B}_{GT} (\omega) = |\langle \omega, \mu ||\tau_{-} \sigma_{\mu}||QRPA \rangle|^2.
\end{equation}
The partial half-lives $t_{1/2}$ were calculated using the relation
\footnotesize{
	\begin{eqnarray}\label{eq35}
		t_{1/2} = \frac{C}{(g_A/g_V)^2f_A(A, Z, E)\mathcal{B}_{GT}(\omega)+f_V(A, Z, E){\mathcal{B}_F(\omega)}},
	\end{eqnarray}
	\normalsize
	where  C (= ${2\pi^3 \hbar^7 ln2}/{g^2_V m^5_ec^4}$) was taken as 6143 $s$ \cite{Har09}. The ratio  $g_A/g_V$ was taken as -1.2694 \cite{Nak10} and $E$ = ($Q$ - $\omega$), where $Q$ represent the Q-value of the reaction and computed from Ref.~\cite{Kon21}. $f_V(A, Z, E)$  and $f_A(A, Z, E)$ are the phase space integrals   for vector transition and axial vector,  respectively. $\mathcal{B}_{GT}$ ($\mathcal{B}_{F}$) is the reduced GT (Fermi) transition probabilities. Calculation of reduced Fermi transitions are relatively simple and can be seen from Ref.~\cite{Hir91}. 
	The terrestrial half-lives of $\beta$-decay were calculated by summing over the inverses of partial half-lives and taking inverse
	\begin{equation}\label{eq36}
		~ ~ ~ ~ ~ ~ ~ ~ ~ ~ ~ ~ ~	T_{1/2} = \left(\sum_{0 \le E_j \le Q} \frac{1}{t_{1/2}}\right)^{-1}.
	\end{equation}
	For details on solution of Eq.~\ref{eq24} we refer to ~\cite{Sta90, Hir93, Mut92, Hir91}.
	
	The stellar $\beta$-decay rate from parent level $n$ to daughter state $m$ was calculated using
	\begin{equation} \label{eq37}
		~ ~ ~ ~ ~ ~ ~ ~ ~ ~ ~ ~ ~	\lambda _{nm}^{\beta} =\ln 2\frac{f_{nm}(T,\rho
			,E_{f})}{(ft)_{nm}}.
	\end{equation}
	The  $(ft)_{nm}$ values are related to the reduced transition probabilities of Fermi and GT transitions
	\begin{equation} \label{eq38}
		~ ~ ~ ~ ~ ~ ~ ~ ~ ~ ~ ~ ~ 	(ft)_{nm} =C/\mathcal{B}^{nm},
	\end{equation}
	where
	\begin{equation} \label{eq39}
		~ ~ ~ ~ ~ ~ ~ ~ ~ ~ ~ ~ ~	\mathcal{B}^{nm}=(g_{A}/g_{V})^{2} \mathcal{B}_{GT}^{nm} + \mathcal{B}_{F}^{nm}.
	\end{equation}
	The reduced Fermi and GT transition probabilities were determined using
	\begin{equation} \label{eq40}
		~ ~ ~ ~ ~ ~ ~ ~ ~ ~ ~ ~ ~	\mathcal{B}_{F}^{nm} = \frac{1}{2J_{n} +1} \langle{m}\parallel\sum\limits_{k}
		\tau_{-}^{k}\parallel {n}\rangle|^{2}
	\end{equation}
	\begin{equation} \label{eq41}
		~ ~ ~ ~ ~ ~ ~ ~ ~ ~ ~ ~ ~	\mathcal{B}_{GT}^{nm} = \frac{1}{2J_{n} +1} \langle{m}\parallel\sum\limits_{k}
		\tau_{-}^{k}\overrightarrow{\sigma}^{k}\parallel {n}\rangle|^{2},
	\end{equation}
	where $\tau_{-}^{k}$ and $\overrightarrow{\sigma}(k)$ are the  isospin lowering and spin operators, respectively. Construction of low-lying excited levels and connecting nuclear matrix elements can be seen from Ref.~\cite{Mut92}.  The phase space  integrals ($f_{nm}$) were calculated using  
	\footnotesize
	\begin{equation} \label{eq42}
		f_{nm} = \int _{1 }^{E_\beta}E_k\sqrt{E_k^{2} -1}(E_{\beta} -E_k)^{2} F(+
		Z, E_k) (1-\mathcal{R}_{-})dE_k,
	\end{equation}
	\normalsize
	where $E_k$ is the electron's kinetic energy including rest mass and we have used natural units ($\hbar=m_{e}=c=1$). The Fermi functions, $F (+Z, E_k)$, were computed as in Ref.~\cite{Gov71}. 
	The total $\beta$-decay energy was calculated using
	\begin{equation} \label{eq43}
		~ ~ ~ ~ ~ ~ ~ ~ ~ ~ ~ ~ ~	E_{\beta} = m_{p} -m_{d} + E_{n} -E_{m},
	\end{equation}
	where $E_{m}$ and $m_{d}$ are the excitation energies and mass of daughter nucleus while, $E_{n}$ and $m_{p}$ represent the corresponding quantities of parent nucleus, respectively. 
	$\mathcal{R}_{-}$ is the electron distribution function given by  
	\begin{equation} \label{eq44}
		~ ~ ~ ~ ~ ~ ~ ~ ~ ~ ~ ~ ~	\mathcal{R}_{-} =\left[\exp \left(\frac{E-E_{f} }{k_{\beta}T} \right)+1\right]^{-1},
	\end{equation}
	where $k_{\beta}$ is the Boltzmann constant, $E$ = ($E_k$ - 1) and $E_{f}$   denote the kinetic  and Fermi energy of the electrons, respectively.

	The electron number density corresponding to protons and nuclei is calculated using the relation 
	\begin{equation} \label{eq45}
		\rho Y_{e}=\frac{1}{\pi^2N_A}\left(\frac{m_ec}{\hbar} \right)^3\int_0^\infty(\mathcal{R}_--\mathcal{R}_+)p^2dp.
	\end{equation}
	$N_{A}$ is Avogadro number, $Y_{e}$ is the ratio of electron number to the baryon number and $\rho$ is the baryon density. The positron distribution function was represented by $\mathcal{R}_{+}$  
	\begin{equation} \label {eq46}
		~ ~ ~ ~ ~ ~ ~ ~ ~ ~ ~ ~ ~	\mathcal{R}_{+} =\left[\exp \left(\frac{E+2-E_{f} }{k_{\beta}T} \right)+1\right]^{-1}.
	\end{equation}
	The total stellar  $\beta$-decay rates were calculated using
	\begin{equation} \label{eq47}
		~ ~ ~ ~ ~ ~ ~ ~ ~ ~ ~ ~ ~	\lambda =\sum _{nm}P_{n} \lambda _{nm}^{\beta},
	\end{equation}
	where $P_n$ is the occupation probability of parent state computed using the Boltzmann distribution. The summation over initial and final states were performed until  convergence was obtained.

\section{Results and discussion}
We first computed the ground-state deformation parameters $\beta_2$ using the RHB formalism with the two interactions, DD-PC1 and DD-ME2. Later on, we used these values as an input parameter in the pn-QRPA model to investigate the $\beta$-decay properties  of 55 neutron-rich nuclei, namely $^{134-139}$Sn, $^{148-153}$Ba, $^{137-144}$Te, $^{145-151}$Cs, $^{134-142}$Sb $^{140-146}$I,  $^{151-155}$La and $^{142-148}$Xe.  

To calculate ground-state deformation parameter ($\beta_2$) values of the selected nuclei, axially symmetric calculations were performed by following the prescription of Ref.~\cite{nik14}. For analysis of deformation parameter, potential energy curves (PECs) of each nucleus was obtained by using quadrupole moment constrained RHB calculations. To obtain each PEC, as a function of $\beta_2$, the ground-state binding energy of nuclei was calculated for fixed $\beta_2$ value. For this purpose, the expectation value of quadrupole moment $\langle Q_2\rangle$ was constrained for a given value $\mu_2$ in the expectation value of the Hamiltonian
\begin{equation}  
	\langle H' \rangle=\langle H\rangle + C_\mu (\langle Q_2\rangle-\mu_2)^2, \label{constr} 
\end{equation}
where $C_\mu$ is the constraint multiplier and

\begin{equation}
	\langle Q_2 \rangle =\frac{3}{\sqrt{5\pi}}Ar^2\beta_2, \label{quad3}
\end{equation} 
where $r = R_0A^{1/3}$ ($R_0 = 1.2$ fm) and $A$ is the mass number of the considered nuclei. 

In Figs. (\ref{Fig. 1}-\ref{Fig. 4}), the calculated PECs of $^{134-139}$Sn,  $^{137-144}$Te, $^{142-148}$Xe and $^{148-153}$Ba, using DD-ME2 and DD-PC1 functionals, are shown. For obtaining the PECs,  lowest binding energy was taken as reference in each PEC calculation. As can be seen in Fig. (\ref{Fig. 1}), the minima are located around $\beta_2=0$ for $^{134-138}$Sn.  Both interactions predict ground-state shape of Sn isotopes to be spherical. In the case of $^{139}$Sn, the minima is located slightly on the right side of $\beta_2=0$ value which may hint towards gradual transition to prolate shape configuration. From the PECs of $^{137-144}$Te, $^{142-148}$Xe and $^{148-153}$Ba, prolate shape configurations are inferred because the minima of each PEC is located on the positive $\beta_2$ values. The calculated ground-state $\beta_2$ values of $^{137-144}$Te, $^{142-148}$Xe and $^{148-153}$Ba with DD-PC1 and DD-ME2 functionals are located around $\beta_2=0.115-0.195$, $\beta_2=0.181-0.233$ and $\beta_2=0.217-0.304$, respectively. It should be noted that similar PECs in Figs. (\ref{Fig. 1}-\ref{Fig. 4}) are obtained with both density-dependent functionals. Prolate shape configurations were also computed for odd-Z nuclei: $^{145-151}$Cs, $^{134-142}$Sb $^{140-146}$I and $^{151-155}$La using RHB formalism with both DD-ME2 and DD-PC1 functionals. For space consideration their PECs are not shown here. The list of ground-state $\beta_2$ values of the considered nuclei obtained from RHB calculations can be seen in Table.~\ref{Tab1}.

The $\beta$-decay half-lives were calculated using three different deformations (FRDM, DD-PC1 and DD-ME2). Our model calculated the branching ratios, GT strength distributions and partial half-lives. We first compared our calculated GT spectra with experimental values to validate our nuclear model. After validation of the chosen model,  we proceeded to calculate stellar $\beta$-decay rates at selected temperature (3, 10, 30) GK and densities ($10^3, 10^7$, $10^{11}$)~g/cm$^3$ using these three deformations.

Table~\ref{Tab1} shows the computed deformation values of selected neutron-rich nuclei using the FRDM and RHB (DD-PC1 and DD-ME2 interactions) calculations. The change in deformation ${\Delta{\beta}_2}$ (Column 6) is  defined as the difference between the maximum and minimum computed absolute $\beta_{2}$ values. The percentage change in half-lives ${\Delta T_{1/2}{\%}}$ (Column 7) is defined as the percentage change in maximum and minimum values of the calculated $\beta$-decay half-life. It was noted that the half-lives of nuclei decreased as the deformation values increased. A similar conclusion was drawn in an earlier investigation using the same model~\cite{Hir93}. The fragmentation of calculated GT strength increased with $\beta_2$ values. Moreover, a significant strength shifted to low-lying states with rising values of $\beta_2$.  Table~\ref{Tab1} shows a maximum  ${\Delta{\beta}_2}$ of 0.13 and a maximum   ${\Delta T_{1/2}{\%}}$ of around 3 orders of magnitude. In majority of the cases, we found a direct correlation between ${\Delta T_{1/2}{\%}}$ and ${\Delta{\beta}_2}$. In the case of $^{143}$Xe, the  half-lives changed by more than 400\%, even though ${\Delta{\beta}_2}$ was not substantial. This is due to fact that employing the DD-PC1 deformation,  resulted in low-lying transition at 0.37 MeV with a corresponding branching ratio of 70.76. The calculated half-life was 148.44 $ms$. Using the FRDM deformation, the corresponding transition shifted to 0.52 MeV with a branching ratio 58.54, yielding a half-life value of 750.99 $ms$. The measured half-life of $^{143}$Xe is 519 $ms$.The half-life of $^{138}$Sb ($^{139}$Sb) was calculated to be 2207.66 (119.19) $ms$ using the FRDM and 260.42 (13.35) $ms$ using the RHB (DD-PC1) calculation. The measured half-life of $^{138}$Sb and $^{139}$Sb are 326 $ms$ and 182 $ms$, respectively. ${\Delta T_{1/2}}$  increased to more than 700$\%$ due to the big difference between the calculated half-lives. The calculated half-lives using DD-PC1 is much smaller than those calculated using the FRDM. This is because of low-lying   transitions calculated using the DD-PC1 interaction. We next show our model performance in reproducing the low-lying GT transitions of the selected neutron-rich nuclei. Fig.~\ref{Fig. GT} shows the comparison of our calculated GT strength distributions of $^{135}$Sn, $^{137,140}$Te  and  $^{148}$Cs with the measured data. In the legend, pn-QRPA-0, pn-QRPA-1 and pn-QRPA-2 represent our model calculations employing deformations from FRDM, DD-PC1 and DD-ME2 functionals, respectively. The experimental data were taken from Refs.~\cite{She05,Si22,Moo17, Lic18}, respectively. The measured data was reported up to  excitation energies (2.461, 2.413, 1.786, 2.557) MeV for $^{135}$Sn, $^{137,140}$Te and $^{148}$Cs, respectively. Calculated GT transitions are shown up to 3 MeV in daughter. For the case of $^{135}$Sn and $^{140}$Te, our model reproduces well the low-lying transitions.  However, in $^{137}$Te and $^{148}$Cs, the low-lying transitions are missing which we take as a shortcoming of our model. Overall, a good agreement between measured and calculated GT strength distributions is noted  for low-lying transitions validating the choice of given nuclear model in our investigation. Our model incorporated residual correlations among nucleons through one-particle one-hole (1p-1h) excitations within a spacious multi-7$\hbar \omega$ model space.  Inclusion of 2p-2h or higher excitations may shift the strength to low-lying states in daughter. It is further noted that the comparison with measured data is best when we used the FRDM calculated deformations in our model (pn-QRPA-0).

Figs.~(\ref{Fig. HL1}-\ref{Fig. HL2}) show ratios of calculated to measured half-lives verses mass number for selected 55 neutron-rich nuclei. Here onward, pn-QRPA-0, pn-QRPA-1 and pn-QRPA-2 represent the pn-QRPA calculated half-lives employing deformations from FRDM, DD-PC1 and DD-ME2 functionals, respectively. The half-lives computed by gross theory~\cite{Wu20} is denoted by GrTh. The gross theory depends on the summation rule of the $\beta$-decay strength function and employs a statistical technique to analyze transitions to final nuclear states~\cite {Tac90}. Fig.~\ref{Fig. HL1} displays the ratios of theoretical to experimental half-lives of  $^{(151-155)}$La, $^{(140-146)}$I, $^{(145-151)}$Cs and $^{(148-153)}$Ba. The pn-QRPA-0  calculation resulted in the best prediction. Similarly, Fig.~\ref{Fig. HL2} shows the ratio of calculated to measured half-lives for $^{(142-148)}$Xe, $^{(137-144)}$Te, $^{(134-139)}$Sn and $^{(134-142)}$Sb. Table~\ref{Tab2} gives a quantitative analysis of the calculations shown in Figs.~(\ref{Fig. HL1}-\ref{Fig. HL2}). In this table, $N$ represents the number of nuclei considered in our investigation, $n$ is the number of calculated half-lives reproduced under the conditions stated in the first column and $\bar{r}$ is the mean deviation given by 
\begin{equation}\label{eq53}
	~ ~ ~ ~ ~ ~ ~ ~ ~ ~ ~ ~ ~ ~ ~ 	\bar{r} = \frac{\sum_{i=1}^{n}r_i}{n},
\end{equation}
where
\begin{equation}\label{eq54}
	r_i = \begin{array}{rcl}
		T^{cal}_{1/2}/T^{exp}_{1/2} 	~~~~~if~~~~~ T^{cal}_{1/2}\geq T^{exp}_{1/2} \\
		T^{exp}_{1/2}/T^{cal}_{1/2} 	~~~~~if~~~~~ T^{exp}_{1/2} > T^{cal}_{1/2}
	\end{array} .
\end{equation}
The pn-QRPA-0 reproduces 100\% and 87.3\% of all measured half-lives within a factor 10 and 2, respectively. The corresponding average deviations are $\bar{r}=1.78$ and $1.35$, respectively. The second best calculation is produced by pn-QRPA-2 where 78.2\% of all calculated half-lives are reproduced within a 2 factor.

Figs.~(\ref{Fig. 08}-\ref{Fig. 15}) depict the GT strength functions, branching ratios and partial half-lives as a function of excitation energy using the pn-QRPA-0, pn-QRPA-1 and pn-QRPA-2 model calculations. Each figure includes three panels: the top and middle panels show the GT strength distributions and branching ratios, respectively, whereas bottom panels represent the partial half-lives as a function of daughter excitation energy. The branching ratios were determined using the relation
\begin{equation}\label{eq55}
	~ ~ ~ ~ ~ ~ ~ ~ ~ ~ ~ ~ ~ ~ ~ ~ ~ ~ ~ ~ I=\frac{T_{1/2}}{t_{1/2}}\times100(\%) \\
\end{equation}
Figs.~(\ref{Fig. 08}-\ref{Fig. 15}) show calculated $\beta$-decay properties for $^{134}$Sn, $^{136}$Sb, $^{139}$Te, $^{140}$I, $^{145}$Xe, $^{146}$Cs, $^{149}$Ba and $^{151}$La, respectively. All GT strengths are shown within the  Q-window. It was noted that for selected deformations, the computed GT strength is well fragmented over the specified energy range. As $\beta_{2}$ increases, there is a tendency for the GT strength to shift to low-lying transitions. Fig. \ref{Fig. 08} shows that the fragmentation of total GT strength increases as the $^{134}$Sn nucleus transitions from spherical to prolate configuration.  Increase in GT strength led to smaller calculated partial half-lives. No substantial change in $\beta$-decay properties was noted when the $\beta_2$ values change marginally.

Table~\ref{Tab3} shows the ratios of computed stellar $\beta$-decay rates for nuclei as the deformation value changes. We selected one nucleus from each isotopic chain to show our results. The stellar rates were computed using the same pn-QRPA model at selected core temperature (3, 10, 30) GK and density (10$^3$, 10$^7$, 10$^{11}$) g/cm$^{3}$ values. The $\beta$-decay rates increase with rising temperature of the stellar core for a given density. This is because the excited state contribution causes the sum to increase at high core temperatures (Eq.~\ref{eq47}). In denser regions of the core, the $\beta$-decay rates start to decrease
significantly due to a sharp reduction in the computed phase space (Eq.~\ref{eq42}). 

The calculated stellar $\beta$-decay rates of $^{138}$Sn decrease by a factor of 13 when using pn-QRPA-1 model as compared to the decay rate calculated using the pn-QRPA-0 model.  The corresponding decrease in calculated half-life is factor 8 when using the pn-QRPA-2 model. For all other cases, except $^{138}$Sn, the change in calculated half-lives with changing deformation values remain within a factor of 2. We further investigated why the impact of changing $\beta_2$ values was so significant for the case of $^{138}$Sn. We noted the magic number $Z$ = 50 for this nucleus. A similar amplified change in stellar $\beta$-decay rates  was noted for other isotopes of Sn, except $^{134}$Sn, used in our investigation. Table~\ref{Tab1} shows  $\Delta \beta_{2}$ value of 0.002 for $^{134}$Sn. The small change in geometrical configuration of $^{134}$Sn resulted in normal deviation of stellar rates within a factor 2.  It was concluded that magic number nucleons react more to changing $\beta_2$ values and substantially alter the stellar rates.

\section{Summary and Conclusions}
In our study, we investigated nuclear structure and $\beta$-decay properties of neutron-rich nuclei in the mass range $134 \leq A \leq 155$. 
The selection of nuclei was motivated by recent half-life measurements at RIBF (RIKEN, Japan). The RHB formalism with DD-PC1 and DD-ME2 functionals was employed to obtain the potential energy curves of each nucleus. Later the ground-state deformation parameters of selected nuclei were deduced. The model-dependent $\beta_{2}$ values were taken as a free parameter in the pn-QRPA model to  investigate $\beta$-decay properties of selected neutron-rich nuclei. GT strength values, branching ratios and half-lives changed significantly with changing $\beta_{2}$ values. 
Fragmentation of the GT strength was noted to increase with the $\beta_{2}$ values. The FRDM computed deformation resulted in best predictive power of the pn-QRPA calculated half-lives followed by the DD-ME2 functional. The terrestrial half-lives changed up to three orders of magnitude as the geometrical configurations of selected nuclei switched. The stellar $\beta$-decay rates changed, at the most, by a factor of two with changing $\beta_{2}$ values. However, for $^{138}$Sn, the half-live changed by more than one order of magnitude. This large deviation in calculated half-life values was also confirmed for other isotopes of magic number Sn nucleus as the nuclear shape changed.  It is hoped that the current investigation will prove useful for a better modeling of nucleosynthesis calculations.

\vspace{0.5in} \textbf{Acknowledgment}:  J.-U. Nabi and A. Mehmood would like to acknowledge financial support of the Higher Education Commission Pakistan through project number 20-15394/NRPU/R\&D/HEC/2021.\\ 
The authors thank Dr. Asim Ullah for useful discussion throughout the course of this study.

\begin{table}[ht]
	\centering
	\caption{Nuclear deformation values ($\beta_{2}$) calculated using FRDM \cite{Mol16} and density-dependent functional (this work).} \label{Tab1}
	\addtolength{\tabcolsep}{1pt}
	\scalebox{0.65}{
		\begin{tabular}{c c c c c c c}
			\hline 
			$\mathbf{A}$ & \textbf{Nuclei} & $\mathbf{\beta_{2}^{FRDM}}$ & $\mathbf{\beta_{2}^{DD-PC1}}$ & $\mathbf{\beta_{2}^{DD-ME2}}$  &  ${\Delta}{\beta}_2$ & ${\Delta T_{1/2}{\%}}$ \\ \hline 
			134 & Sb & -0.021 & 0.002 & 0.000 & 0.021 & 28.7 \\  
			135 & Sb & -0.021 & 0.004 & 0.001 & 0.020 & 25.6 \\  
			136 & Sb & 0.021 & 0.012 & 0.003 & 0.018 & 61.6 \\  
			137 & Sb & -0.021 & 0.106 & 0.015 & 0.091 & 25.9 \\  
			138 & Sb & -0.032 & 0.125 & 0.118 & 0.093 & 747.7 \\  
			139 & Sb & 0.054 & 0.141 & 0.133 & 0.087 & 792.8 \\  
			140 & Sb & 0.065 & 0.150 & 0.142 & 0.085 & 88.1 \\  
			141 & Sb & 0.097 & 0.158 & 0.147 & 0.061 & 78.0 \\  
			142 & Sb & 0.107 & 0.166 & 0.150 & 0.059 & 64.0 \\  
			148 & Ba & 0.228 & 0.224 & 0.217 & 0.010 & 1.2 \\  
			149 & Ba & 0.239 & 0.240 & 0.227 & 0.013 & 0.8 \\  
			150 & Ba & 0.237 & 0.272 & 0.243 & 0.035 & 22.1 \\  
			151 & Ba & 0.249 & 0.287 & 0.294 & 0.045 & 16.3 \\  
			152 & Ba & 0.249 & 0.296 & 0.298 & 0.049 & 12.0 \\  
			153 & Ba & 0.259 & 0.304 & 0.303 & 0.045 & 14.7 \\  
			145 & Cs & 0.163 & 0.202 & 0.199 & 0.039 & 82.1 \\  
			146 & Cs & 0.174 & 0.209 & 0.205 & 0.035 & 19.8 \\  
			147 & Cs & 0.206 & 0.219 & 0.211 & 0.013 & 62.9 \\  
			148 & Cs & 0.216 & 0.230 & 0.219 & 0.014 & 7.7 \\  
			149 & Cs & 0.227 & 0.244 & 0.231 & 0.017 & 5.6 \\  
			150 & Cs & 0.237 & 0.259 & 0.245 & 0.022 & 2.2 \\  
			151 & Cs & 0.226 & 0.273 & 0.263 & 0.047 & 14.7 \\  
			140 & I & 0.107 & 0.158 & 0.157 & 0.051 & 1.8 \\  
			141 & I & 0.118 & 0.174 & 0.173 & 0.056 & 54.9 \\  
			142 & I & 0.141 & 0.180 & 0.179 & 0.039 & 3.9 \\  
			143 & I & 0.151 & 0.187 & 0.184 & 0.036 & 117.5 \\  
			144 & I & 0.152 & 0.194 & 0.188 & 0.042 & 42.9 \\  
			145 & I & 0.172 & 0.203 & 0.193 & 0.031 & 11.5 \\  
			146 & I & 0.194 & 0.215 & 0.199 & 0.021 & 133.3 \\  
			151 & La & 0.249 & 0.311 & 0.314 & 0.065 & 14.0 \\  
			152 & La & 0.260 & 0.312 & 0.314 & 0.054 & 15.6 \\  
			153 & La & 0.259 & 0.315 & 0.316 & 0.057 & 3.1 \\  
			154 & La & 0.259 & 0.321 & 0.319 & 0.062 & 7.0 \\  
			155 & La & 0.271 & 0.328 & 0.326 & 0.057 & 10.4 \\  
			137 & Te & 0.011 & 0.116 & 0.115 & 0.105 & 190.8 \\  
			138 & Te & 0.000 & 0.134 & 0.133 & 0.134 & 18.1 \\  
			139 & Te & 0.087 & 0.150 & 0.149 & 0.063 & 86.3 \\  
			140 & Te & 0.097 & 0.165 & 0.164 & 0.068 & 66.3 \\  
			141 & Te & 0.118 & 0.172 & 0.170 & 0.054 & 56.7 \\  
			142 & Te & 0.118 & 0.179 & 0.175 & 0.061 & 8.2 \\  
			143 & Te & 0.150 & 0.186 & 0.178 & 0.036 & 6.4 \\  
			144 & Te & 0.161 & 0.195 & 0.182 & 0.034 & 1.9 \\  
			134 & Sn & 0.000 & 0.002 & 0.001 & 0.002 & 8.2 \\  
			135 & Sn & 0.000 & 0.003 & 0.002 & 0.003 & 1.7 \\  
			136 & Sn & 0.000 & 0.005 & 0.003 & 0.005 & 3.9 \\  
			137 & Sn & 0.000 & 0.008 & 0.004 & 0.008 & 2.8 \\  
			138 & Sn & 0.000 & 0.020 & 0.006 & 0.020 & 4.5 \\  
			139 & Sn & 0.000 & 0.101 & 0.009 & 0.101 & 36.0 \\  
			142 & Xe & 0.141 & 0.181 & 0.181 & 0.040 & 24.4 \\  
			143 & Xe & 0.152 & 0.188 & 0.186 & 0.036 & 405.9 \\  
			144 & Xe & 0.162 & 0.194 & 0.191 & 0.032 & 2.3 \\  
			145 & Xe & 0.174 & 0.202 & 0.197 & 0.028 & 21.2 \\  
			146 & Xe & 0.194 & 0.210 & 0.202 & 0.016 &  6.0 \\ 
			147 & Xe & 0.216 & 0.221 & 0.209 & 0.012 &  4.7 \\ 
			148 & Xe & 0.215 & 0.233 & 0.217 & 0.018 &  8.7 \\   \hline
	\end{tabular}}
\end{table}

\begin{table}[ht]
	\centering
	\caption{Accuracy of the pn-QRPA model (using different deformations) and gross theory compared to experimental data. For explanation of variables see text.} 
	\label{Tab2}
	\addtolength{\tabcolsep}{1pt}
	\scalebox{0.8}{
		\begin{tabular}{l l |l l l |l l l |l l l| l l l}
			\hline
			\textbf{Condition} & \textbf{N} & \multicolumn{3}{ l| }{\textbf{\hspace{0.8cm}pn-QRPA-0}} &\multicolumn{3}{ l| }{\textbf{\hspace{0.8cm}pn-QRPA-1}} & \multicolumn{3}{ l| }{\textbf{\hspace{0.8cm}pn-QRPA-2}} &\multicolumn{3}{ l }{\textbf{\hspace{0.8cm}GrTh}}\\ \hline
			~ & ~ & n & n(\%) & ${\bar{r}}$\hspace{0.8cm} & n & n(\%) & ${\bar{r}}$\hspace{0.8cm} & n & n(\%) & ${\bar{r}}$\hspace{0.8cm} & n & n(\%) & ${\bar{r}}$ \\ 
			${r_i}$ \ ${\leq 10}$~ & 55\hspace{0.8cm} & 55 & 100.00 & 1.78 & 54 & 98.18 & 1.96 & 54 & 98.18 & 1.88 & 55 & 100.00 & 2.08 \\ 
			${r_i}$  \ ${\leq2}$~ & 55 & 48 & \hspace{0.05cm} 87.27 & 1.35 & 40 & 72.72 & 1.39 & 43 & 78.18 & 1.39 & 31 & \hspace{0.05cm} 56.36 & 1.42 \\\hline
	\end{tabular}}
\end{table}

\begin{table}[ht]
	\centering
	\caption{Ratios of calculated stellar $\beta$-decay rates pn-QRPA-0 to pn-QRPA-1 and pn-QRPA-0 to pn-QRPA-2 for selected nuclei as a function of core temperatures at densities (given in units of g/cm$^3$). Temperatures ($T_9$) are given in units of $10^9$K. } \label{Tab3}
	\addtolength{\tabcolsep}{1pt}
	\scalebox{0.7}{
		\begin{tabular}{ c|c|c c c|c c c }
			\hline
			~ & ~ & \textbf{$\rho Y_{e}=10^3$} & $\rho Y_{e}=10^7$ & $\rho Y_{e}=10^{11}$ & $\rho Y_{e}=10^3$ & $\rho Y_{e}=10^7$ & $\rho Y_{e}=10^{11}$ \\ \hline
			\textbf{Nuclei} & \textbf{$T_9$} & ~ & $\lambda_{(pn-QRPA-0)}$/$\lambda_{(pn-QRPA-1)}$ & ~ & ~ & $\lambda_{(pn-QRPA-0)}$/$\lambda_{(pn-QRPA-2)}$ & ~ \\ \hline
			
			$^{138}$Sn & 3 & 1.20 & 1.20 & 5.66 & 1.17 & 1.17 & 3.69 \\ 
			~ & 10 & 6.92 & 6.92 & 10.55 & 4.68 & 4.68 & 6.34 \\ 
			~ & 30 & 11.20 & 11.20 & 12.85 & 7.26 & 7.26 & 8.15 \\ \hline
			$^{140}$Sb & 3 & 0.51 & 0.51 & 1.76 & 0.54 & 0.54 & 1.67 \\ 
			~ & 10 & 0.70 & 0.70 & 1.07 & 0.76 & 0.76 & 1.12 \\ 
			~ & 30 & 0.62 & 0.62 & 0.60 & 0.70 & 0.70 & 0.68 \\ \hline
			$^{143}$I & 3 & 0.81 & 0.80 & 1.62 & 0.82 & 0.82 & 1.92 \\ 
			~ & 10 & 0.92 & 0.92 & 0.99 & 0.96 & 0.96 & 0.99 \\ 
			~ & 30 & 0.98 & 0.98 & 0.97 & 1.01 & 1.02 & 0.99 \\ \hline
			$^{143}$Te & 3 & 1.13 & 1.13 & 0.62 & 1.01 & 1.01 & 0.82 \\ 
			~ & 10 & 1.06 & 1.06 & 1.02 & 0.99 & 0.99 & 1.04 \\ 
			~ & 30 & 1.02 & 1.02 & 1.03 & 1.05 & 1.05 & 1.06 \\ \hline 
			$^{147}$Xe & 3 & 1.00 & 1.00 & 0.88 & 1.07 & 1.07 & 1.12 \\ 
			~ & 10 & 1.04 & 1.04 & 0.96 & 0.98 & 0.98 & 0.97 \\ 
			~ & 30 & 1.06 & 1.06 & 1.04 & 0.93 & 0.92 & 0.92 \\ \hline
			$^{150}$Ba & 3 & 1.06 & 1.06 & 0.62 & 1.03 & 1.03 & 0.97 \\ 
			~ & 10 & 0.89 & 0.88 & 0.73 & 0.87 & 0.86 & 0.86 \\ 
			~ & 30 & 0.88 & 0.88 & 0.85 & 0.86 & 0.86 & 0.86 \\ \hline
			$^{151}$Cs & 3 & 0.99 & 0.99 & 0.74 & 1.25 & 1.24 & 1.27 \\ 
			~ & 10 & 0.84 & 0.84 & 0.60 & 1.26 & 1.27 & 1.14 \\ 
			~ & 30 & 0.86 & 0.86 & 0.81 & 1.10 & 1.10 & 1.09 \\ \hline
			$^{154}$La & 3 & 0.95 & 0.96 & 0.79 & 0.92 & 0.93 & 0.70 \\ 
			~ & 10 & 1.04 & 1.04 & 1.01 & 0.87 & 0.87 & 0.84 \\ 
			~ & 30 & 1.01 & 1.01 & 1.03 & 0.78 & 0.78 & 0.78 \\ \hline
	\end{tabular}}
\end{table}
\clearpage

\begin{figure}[ht]
	\centering	
	\includegraphics[width=14cm]{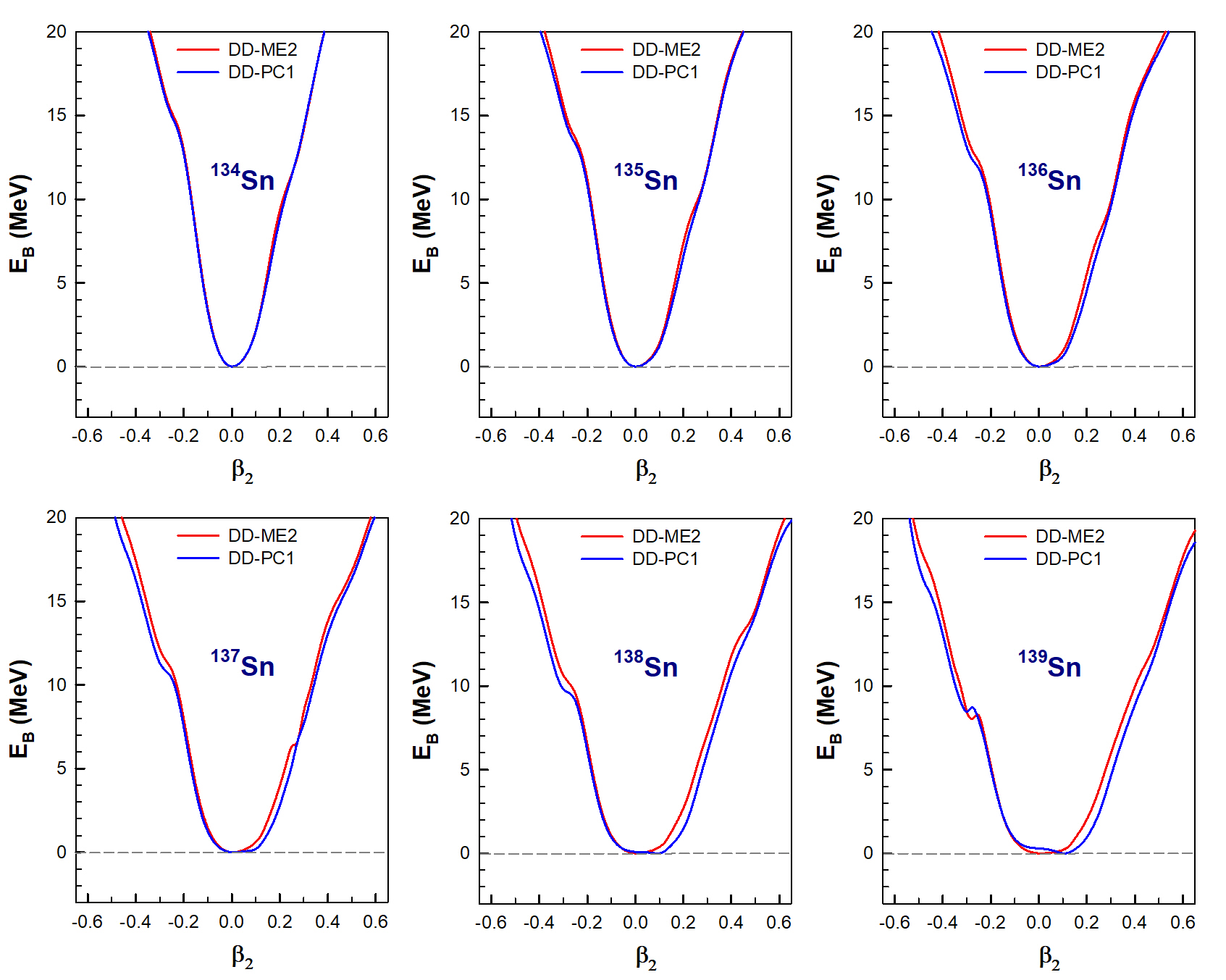}
	\caption{Calculated PECs of $^{134-139}$Sn obtained using DD-ME2 and DD-PC1 functionals.} 
	\label{Fig. 1} 
\end{figure}

\begin{figure}[ht]
	\centering	
	\includegraphics[width=14cm]{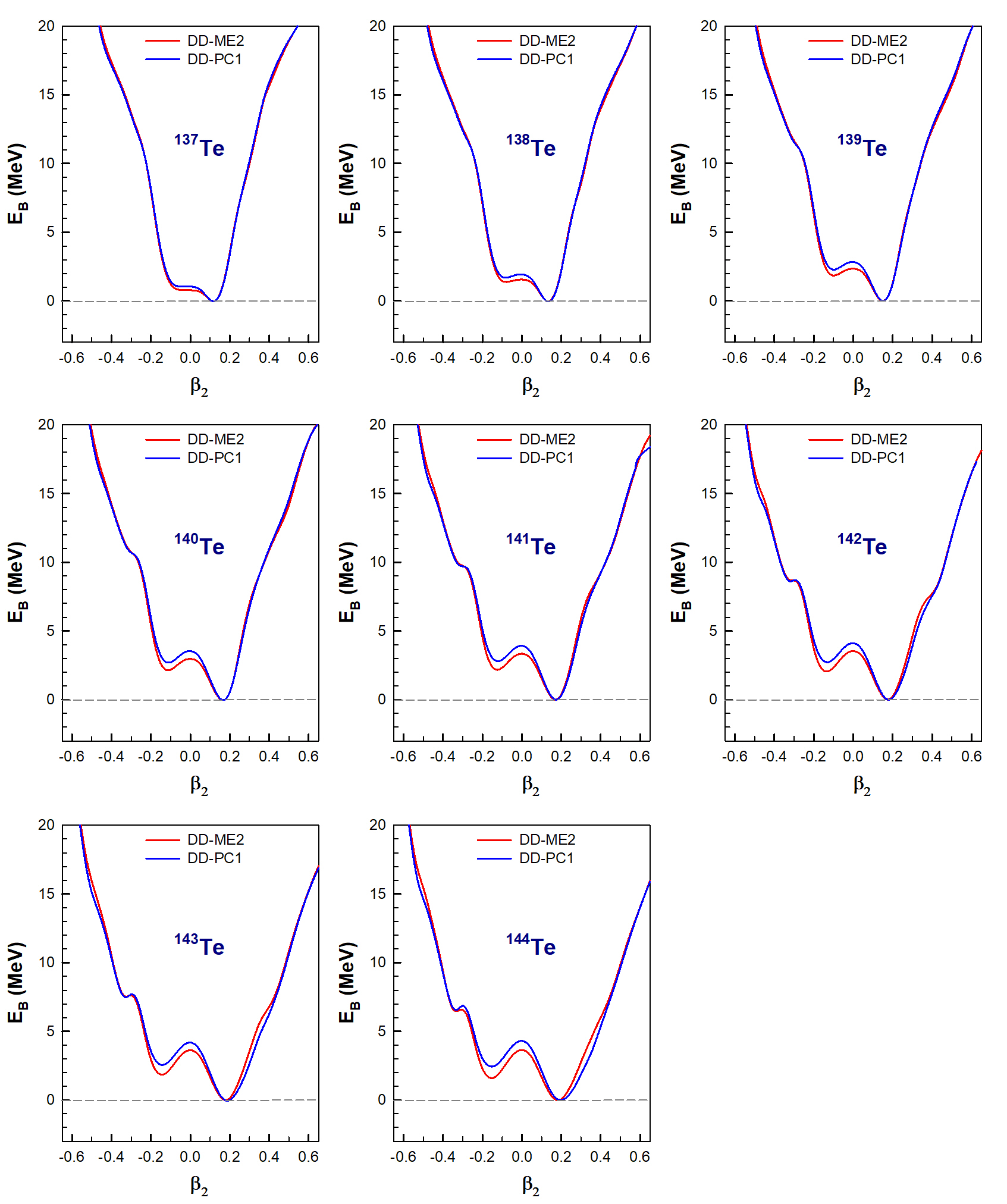}
	\caption{Same as Fig.~\ref{Fig. 1}, but for $^{144-149}$Te.} 
	\label{Fig.2} 
\end{figure}

\begin{figure}[ht]
	\centering	
	\includegraphics[width=14cm]{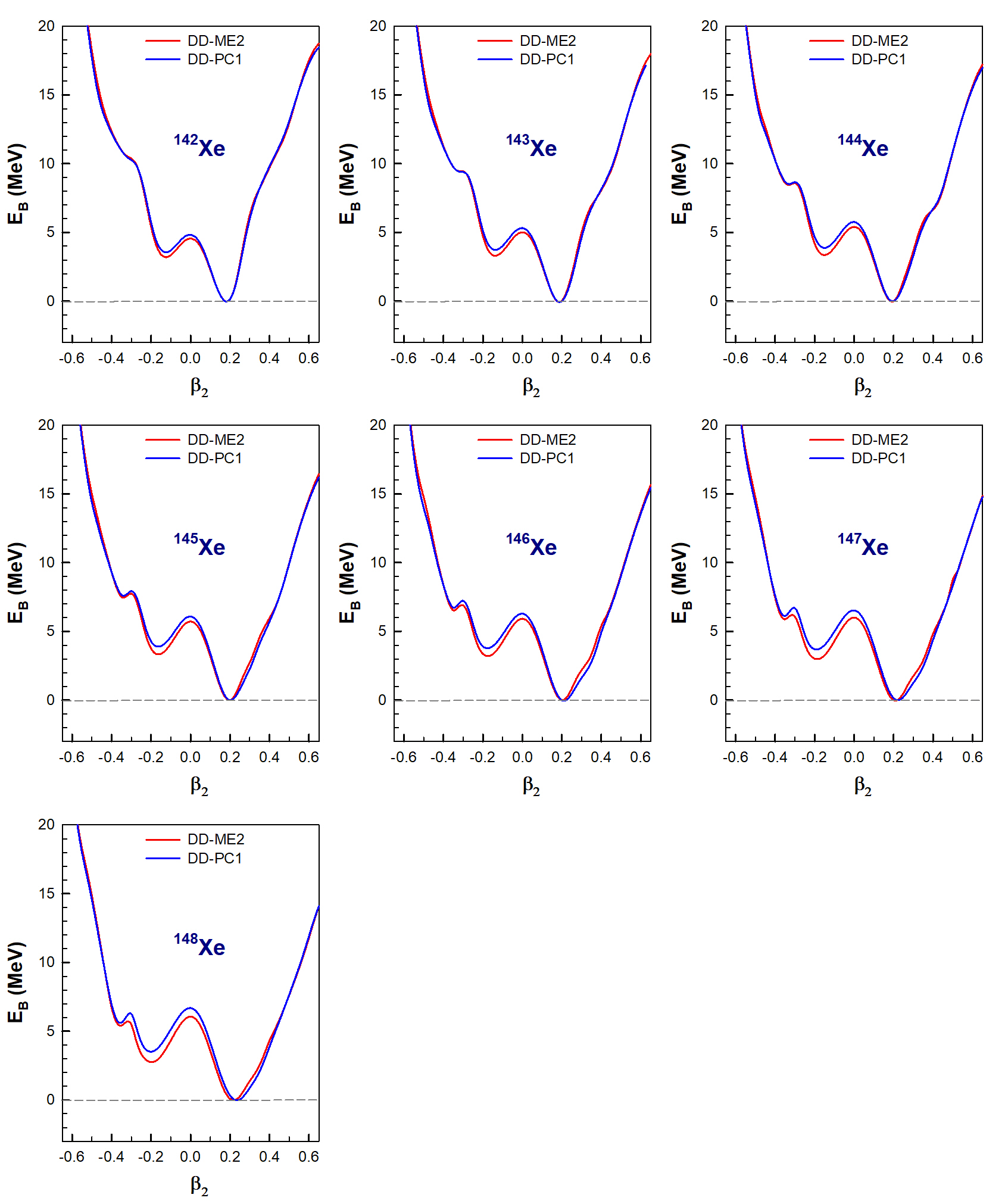}
	\caption{Same as Fig.~\ref{Fig. 1}, but for $^{138-143}$Xe.} 
	\label{Fig. 3} 
\end{figure}

\begin{figure}[ht]
	\centering	
	\includegraphics[width=14cm]{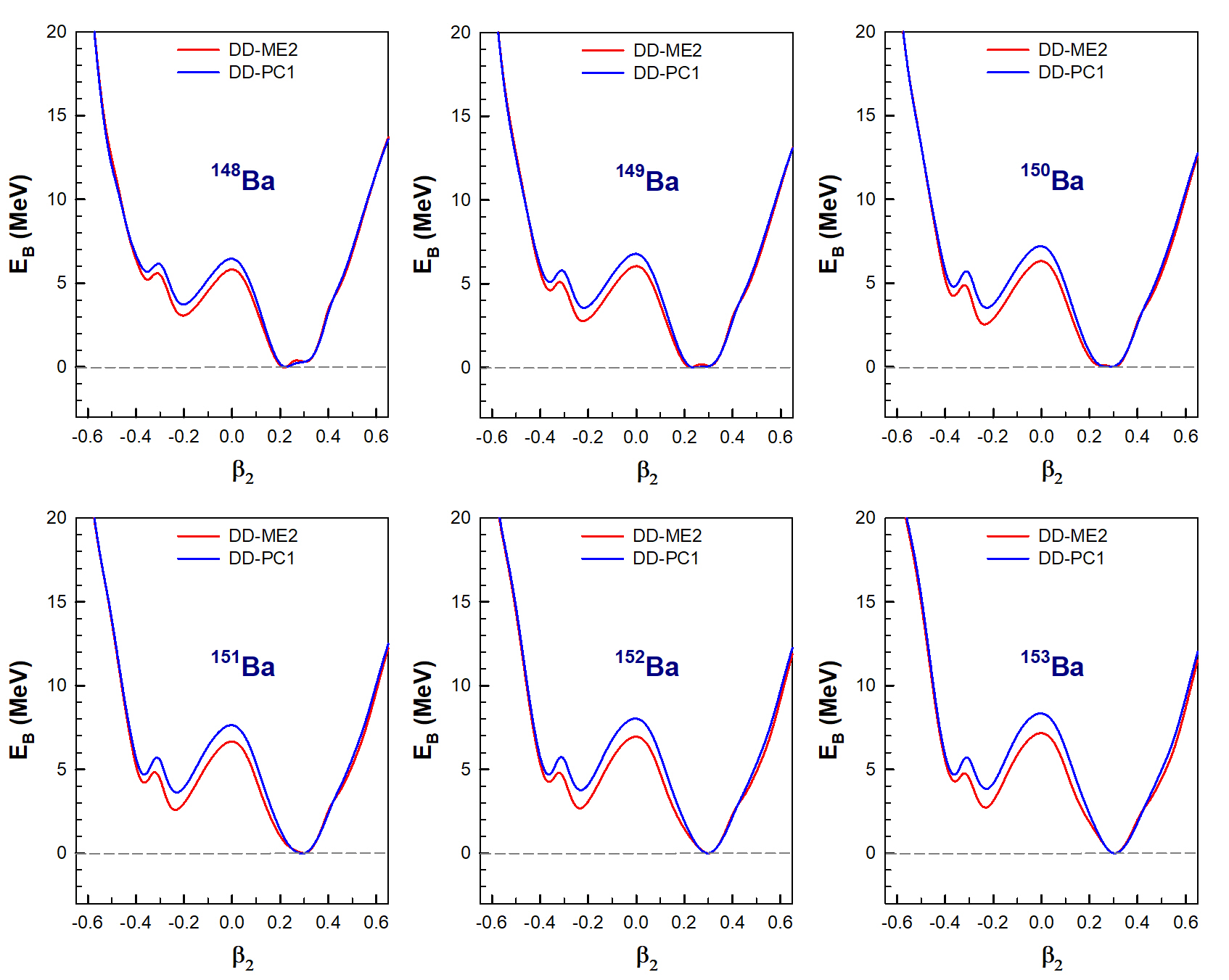}
	\caption{Same as Fig.~\ref{Fig. 1}, but for $^{140-145}$Ba.} 
	\label{Fig. 4} 
\end{figure}

\begin{figure}[ht]
	\centering	
	\includegraphics[height=10cm,width=14cm]{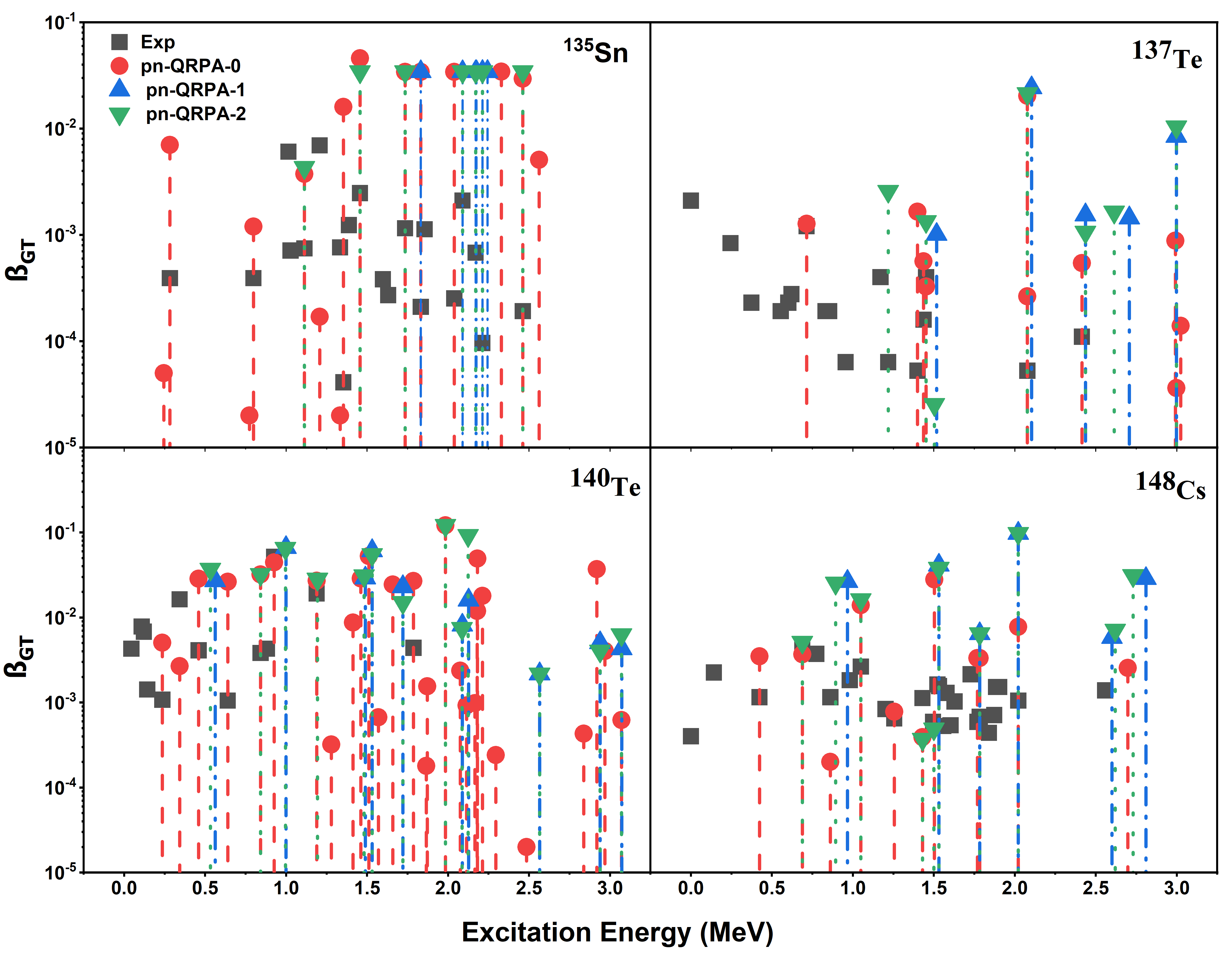}\\
	\caption{Comparison of the pn-QRPA-0, pn-QRPA-1 and pn-QRPA-2 calculated GT strength distributions with experimental data~\cite{She05,Si22,Moo17, Lic18}.}
	\label{Fig. GT} 
\end{figure}

\begin{figure}[ht]
	\centering	
	\includegraphics[height=15cm,width=14cm]{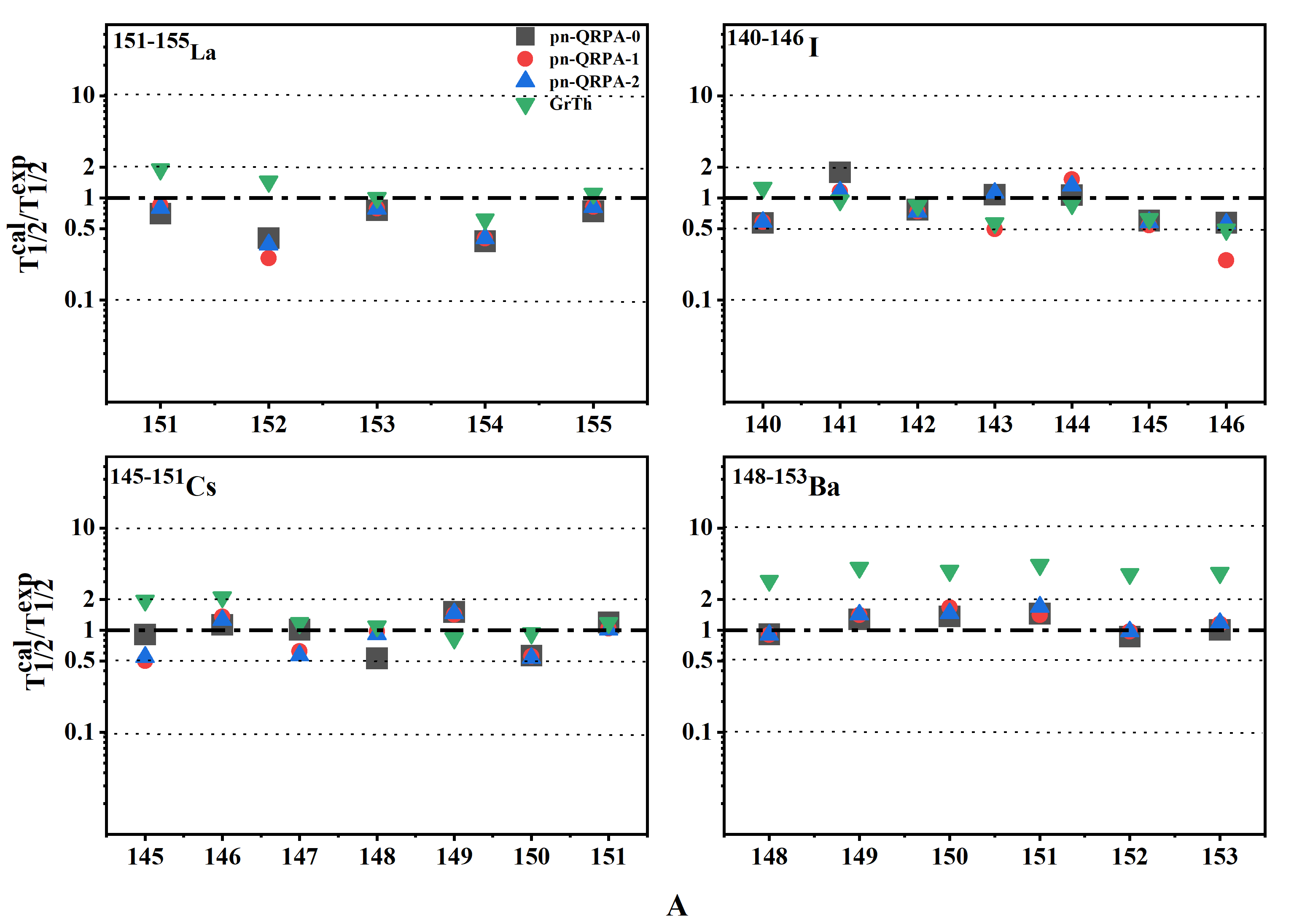}
	\caption{ Ratios of pn-QRPA (using different deformations) and gross theory calculated half-lives of La, Cs, I and Ba isotopes to measured ones. Experimental data was taken from Ref.~\cite{Kon21}.} 
	\label{Fig. HL1} 
\end{figure}

\begin{figure}[ht]
	\centering	
	\includegraphics[height=15cm,width=14cm]{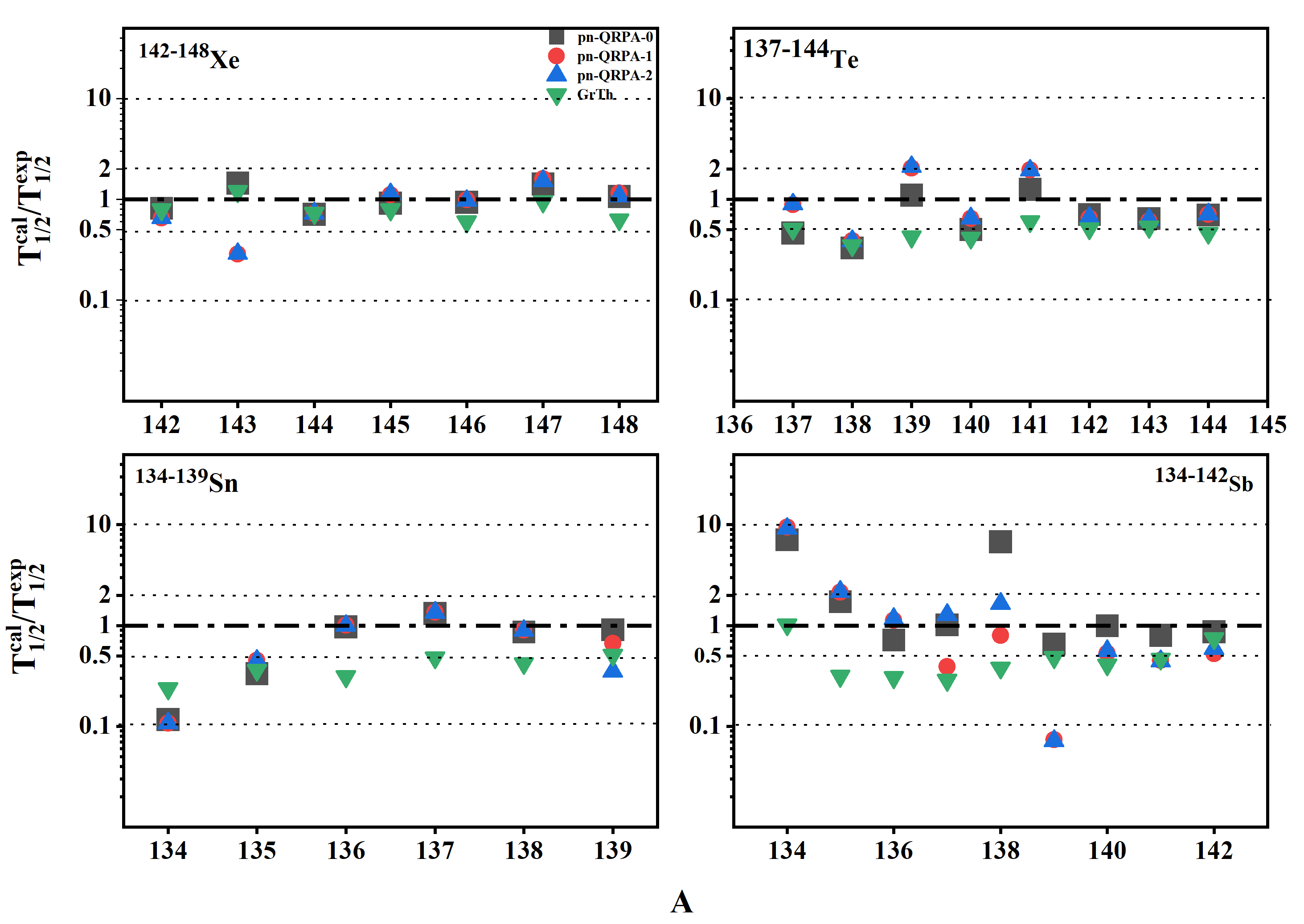}\\
	\caption{Same as Fig.~\ref{Fig. HL1} but for Xe, Te, Sn and Sb isotopes.}
	\label{Fig. HL2}  
\end{figure}

\begin{figure}[ht]
	\centering	
	\includegraphics[height=10cm,width=14cm]{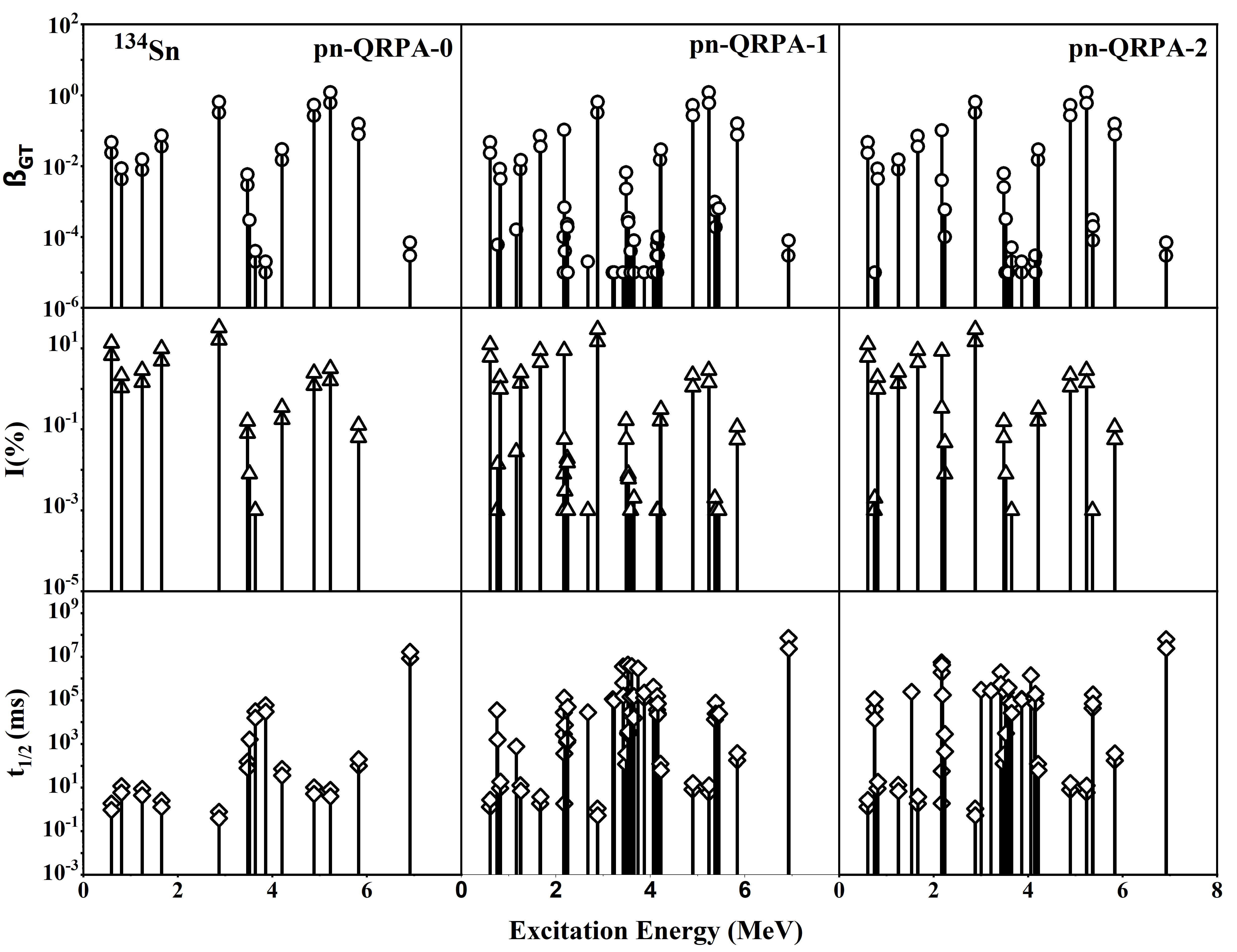}\\
	\caption{Calculated GT strength distributions, branching ratios and partial half-lives of $^{134}$Sn using three model-dependent $\beta_2$ values.}
	\label{Fig. 08} 
\end{figure}

\begin{figure}[ht]
	\centering	
	\includegraphics[height=10cm,width=14cm]{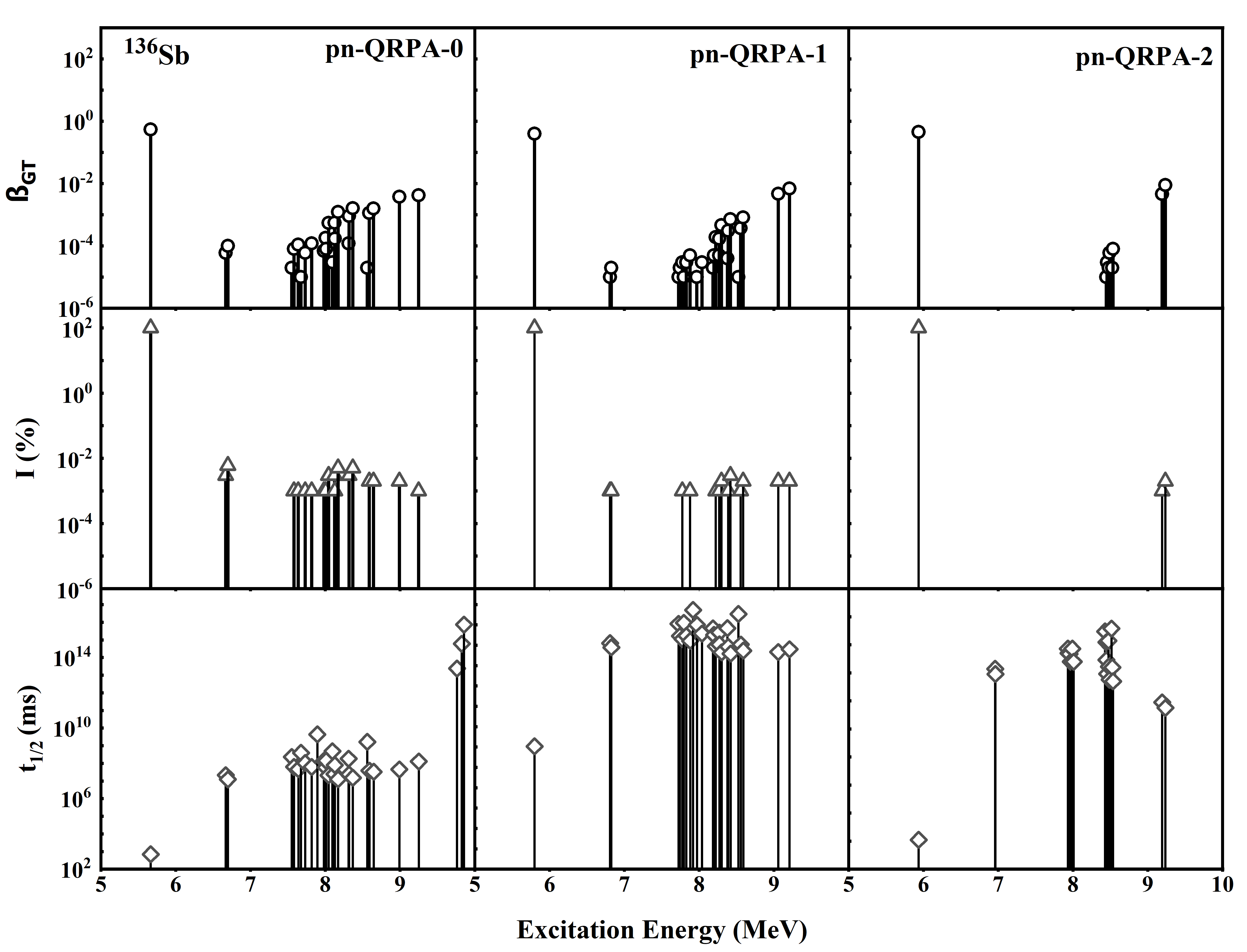}\\
	\caption{Same as Fig.~\ref{Fig. 08} but for $^{136}$Sb.}
	\label{Fig. 09} 
\end{figure}

\begin{figure}[ht]
	\centering	
	\includegraphics[height=10cm,width=14cm]{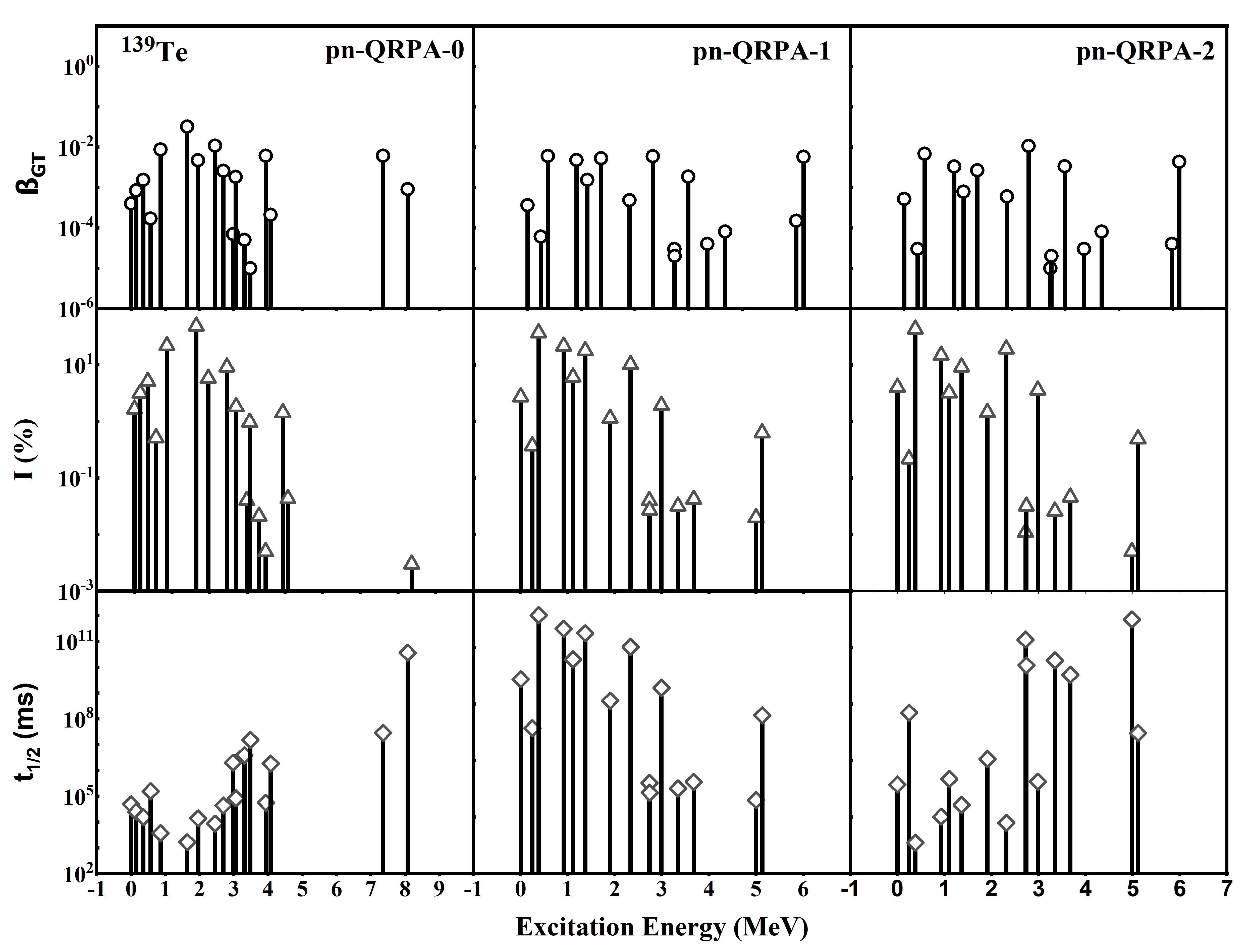}\\
	\caption{Same as Fig.~\ref{Fig. 08} but for $^{139}$Te.}
	\label{Fig. 10} 
\end{figure}

\begin{figure}[ht]
	\centering	
	\includegraphics[height=10cm,width=14cm]{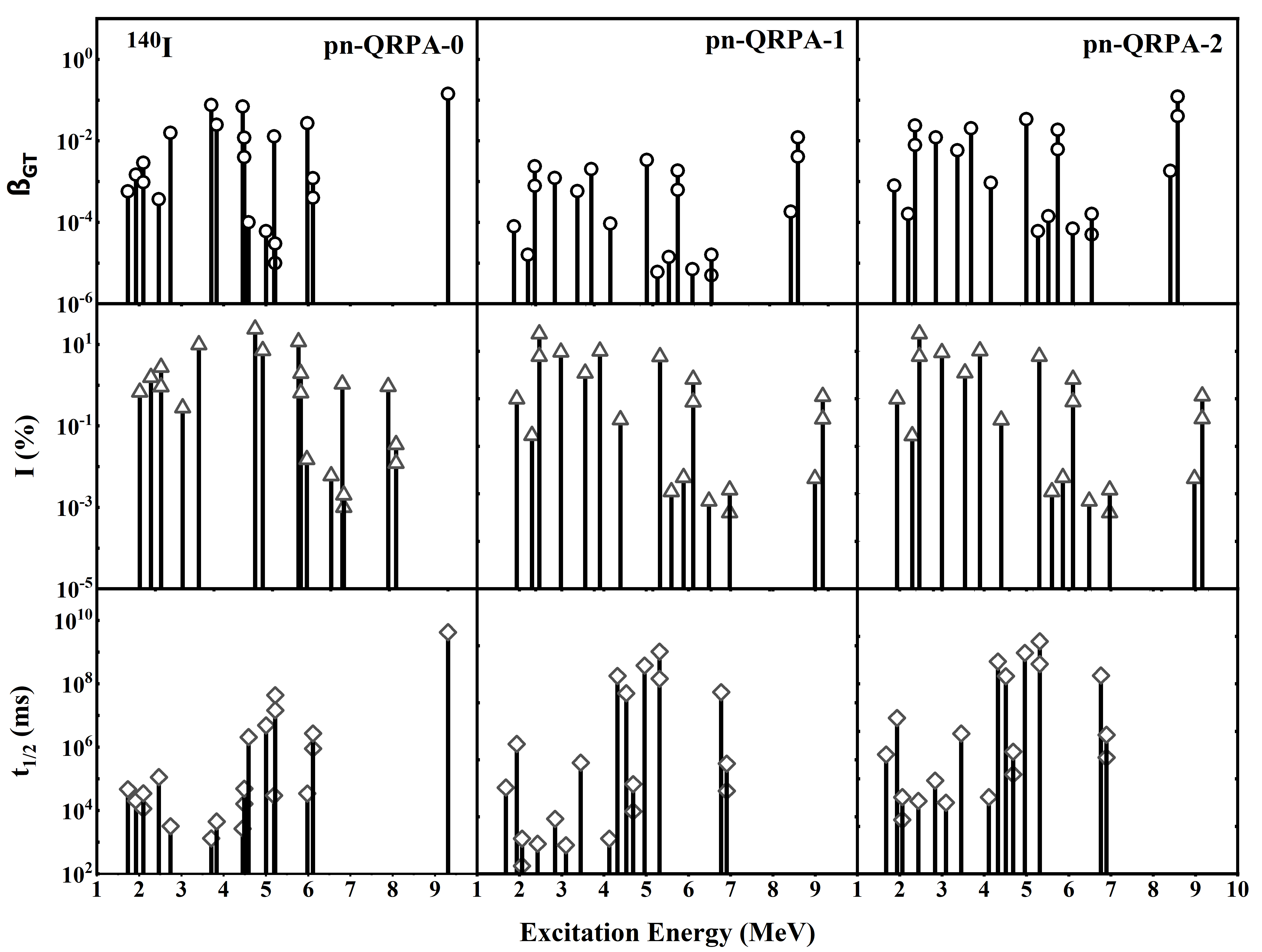}\\
	\caption{Same as Fig.~\ref{Fig. 08} but for $^{140}$I.}
	\label{Fig. 11} 
\end{figure}

\begin{figure}[ht]
	\centering	
	\includegraphics[height=10cm,width=14cm]{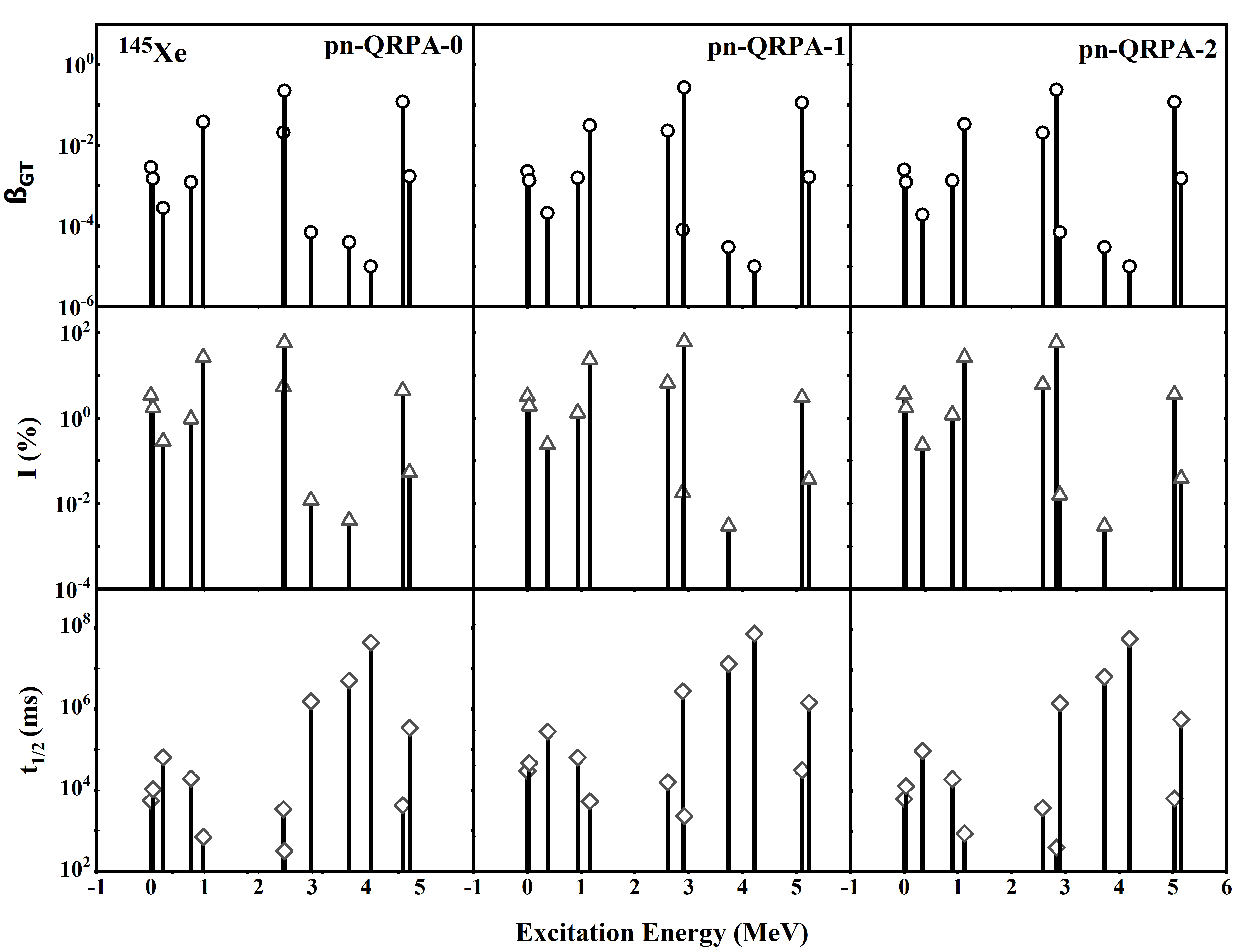}\\
	\caption{Same as Fig.~\ref{Fig. 08} but for $^{145}$Xe.}
	\label{Fig. 12} 
\end{figure}

\begin{figure}[ht]
	\centering	
	\includegraphics[height=10cm,width=14cm]{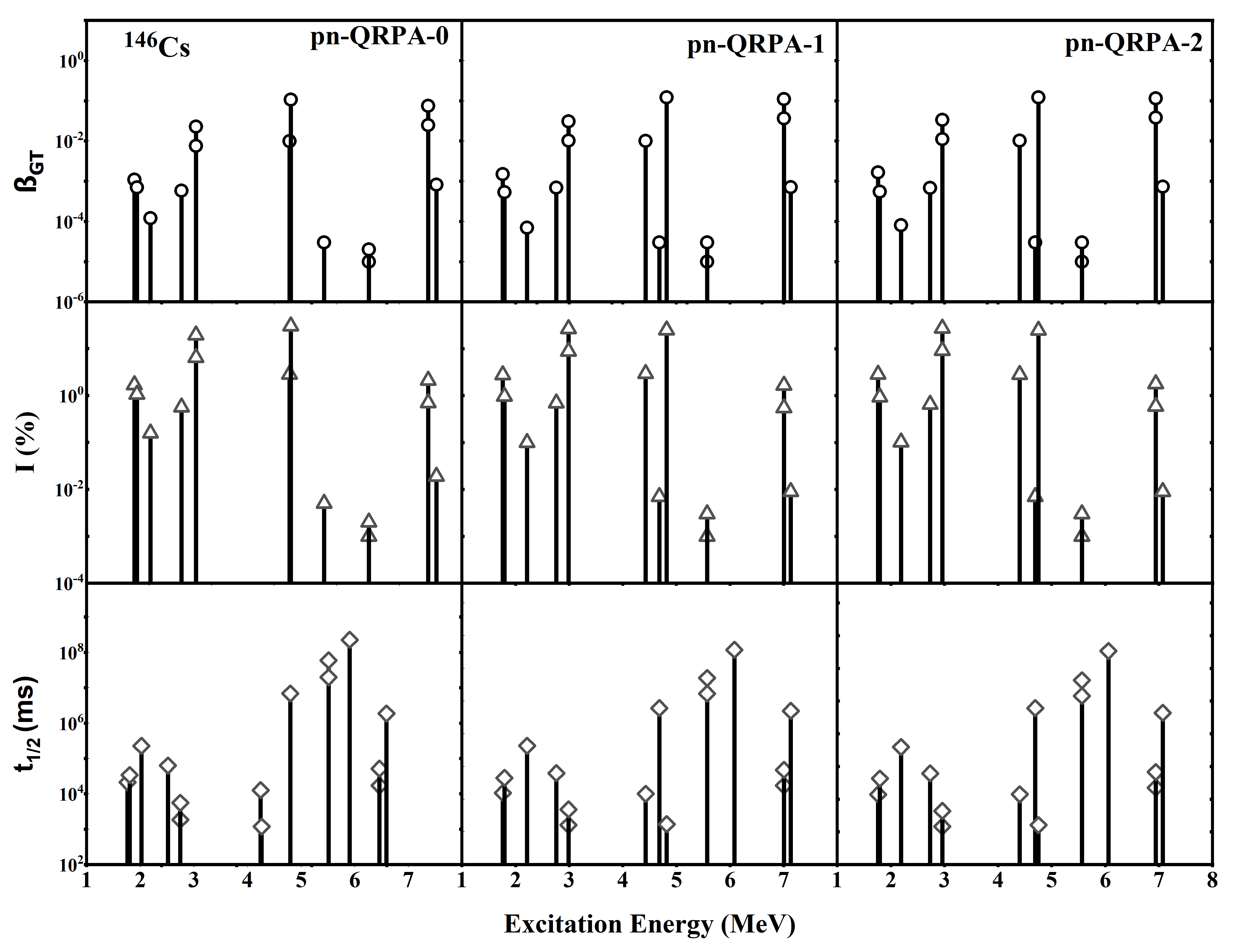}\\
	\caption{Same as Fig.~\ref{Fig. 08} but for $^{146}$Cs.}
	\label{Fig. 13} 
\end{figure}

\begin{figure}[ht]
	\centering	
	\includegraphics[height=10cm,width=14cm]{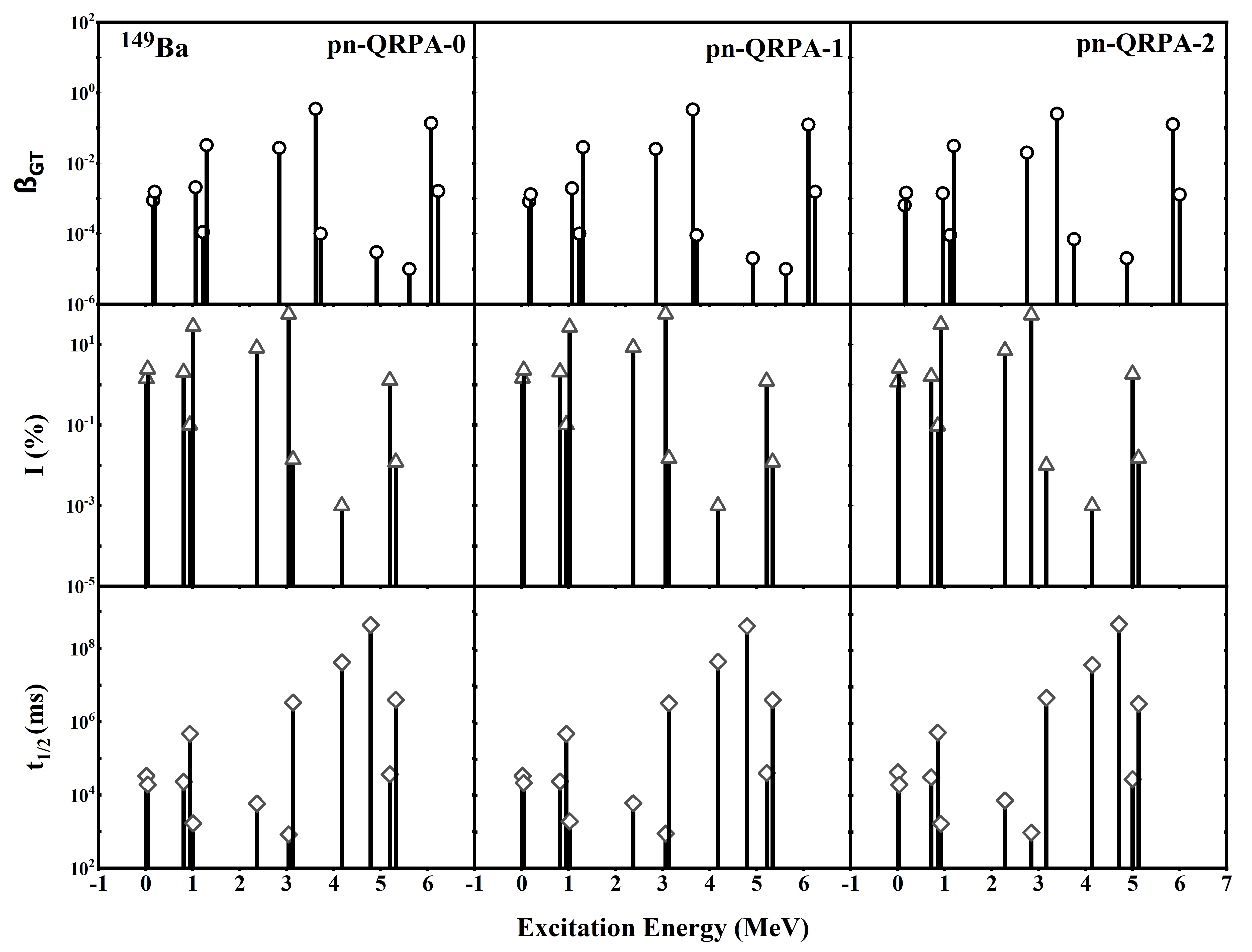}\\
	\caption{Same as Fig.~\ref{Fig. 08} but for $^{149}$Ba.}
	\label{Fig. 14} 
\end{figure}

\begin{figure}[ht]
	\centering	
	\includegraphics[height=10cm,width=14cm]{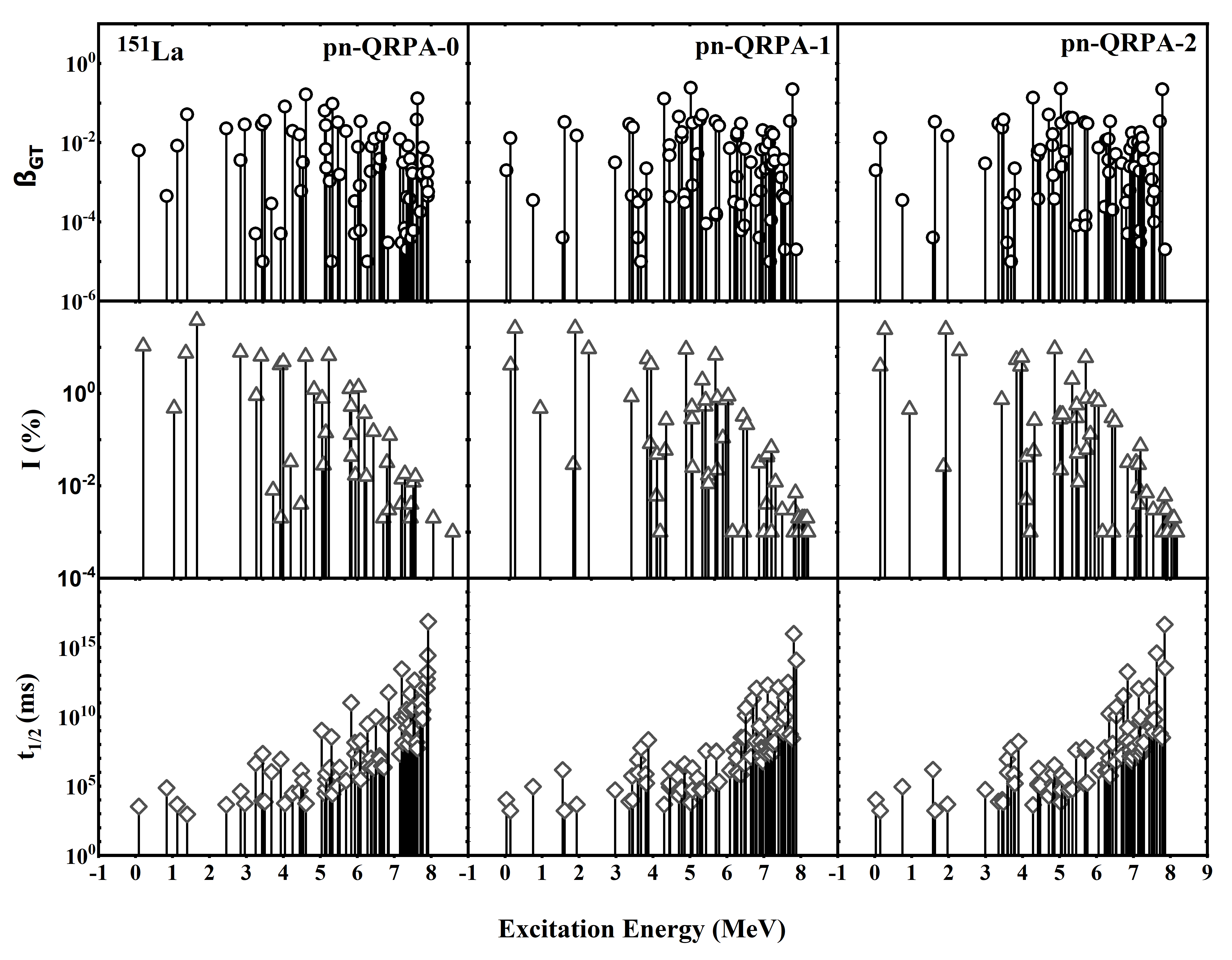}\\
	\caption{Same as Fig.~\ref{Fig. 08} but for $^{151}$La.}
	\label{Fig. 15} 
\end{figure}


\begin{thebibliography}{53}
\expandafter\ifx\csname natexlab\endcsname\relax\def\natexlab#1{#1}\fi
\ifx\xfnm\relax \def\xfnm[#1]{\unskip,\space#1}\fi

\bibitem[{Bayram \& Yilmaz(2013)}]{bayram2013table}
\bibinfo{author}{Bayram, T.},  \& \bibinfo{author}{Yilmaz, A.~H.}
  (\bibinfo{year}{2013}).
\newblock \bibinfo{title}{Table of ground state properties of nuclei in the rmf
  model}.
\newblock {\it \bibinfo{journal}{Mod. Phys. Lett. A.}\/},  {\it
  \bibinfo{volume}{28}\/}\bibinfo{issue}{(16)}, \bibinfo{pages}{1350068}.

\bibitem[{Bender et~al.(2009)Bender, Bennaceur, Duguet, Heenen, Lesinski \&
  Meyer}]{bender2009tensor}
\bibinfo{author}{Bender, M.}, \bibinfo{author}{Bennaceur, K.},
  \bibinfo{author}{Duguet, T.} et~al. (\bibinfo{year}{2009}).
\newblock \bibinfo{title}{Tensor part of the skyrme energy density functional.
  ii. deformation properties of magic and semi-magic nuclei}.
\newblock {\it \bibinfo{journal}{Phys. Rev. C.-Nucl. Phys.}\/},  {\it
  \bibinfo{volume}{80}\/}\bibinfo{issue}{(6)}, \bibinfo{pages}{064302}.

\bibitem[{Boguta \& Bodmer(1977)}]{boguta1977relativistic}
\bibinfo{author}{Boguta, J.},  \& \bibinfo{author}{Bodmer, A.~R.}
  (\bibinfo{year}{1977}).
\newblock \bibinfo{title}{Relativistic calculation of nuclear matter and the
  nuclear surface}.
\newblock {\it \bibinfo{journal}{Nucl. Phys. A.}\/},  {\it
  \bibinfo{volume}{292}\/}\bibinfo{issue}{(3)}, \bibinfo{pages}{413--428}.

\bibitem[{Borzov et~al.(1996)Borzov, Fayans, Kr{\"o}mer \&
  Zawischa}]{borzov1996ground}
\bibinfo{author}{Borzov, I.~N.}, \bibinfo{author}{Fayans, S.~A.},
  \bibinfo{author}{Kr{\"o}mer, E.} et~al. (\bibinfo{year}{1996}).
\newblock \bibinfo{title}{Ground state properties and $\beta$-decay half-lives
  near 132 sn in a self-consistent theory}.
\newblock {\it \bibinfo{journal}{Z. Phys. A: Hadrons Nucl.}\/},  {\it
  \bibinfo{volume}{355}\/}, \bibinfo{pages}{117--127}.

\bibitem[{Burbidge et~al.(1957)Burbidge, Burbidge, Fowler \&
  Hoyle}]{burbidge1957synthesis}
\bibinfo{author}{Burbidge, E.~M.}, \bibinfo{author}{Burbidge, G.~R.},
  \bibinfo{author}{Fowler, W.~A.} et~al. (\bibinfo{year}{1957}).
\newblock \bibinfo{title}{Synthesis of the elements in stars}.
\newblock {\it \bibinfo{journal}{Rev. Mod. Phys.}\/},  {\it
  \bibinfo{volume}{29}\/}\bibinfo{issue}{(4)}, \bibinfo{pages}{547}.

\bibitem[{Cowan et~al.(2021)Cowan, Sneden, Lawler, Aprahamian, A.~Wiescher,
  Langanke, Mart{\'\i}nez-Pinedo \& Thielemann}]{cowan2021origin}
\bibinfo{author}{Cowan, J.~J.}, \bibinfo{author}{Sneden, C.},
  \bibinfo{author}{Lawler, J.~E.} et~al. (\bibinfo{year}{2021}).
\newblock \bibinfo{title}{Origin of the heaviest elements: The rapid
  neutron-capture process}.
\newblock {\it \bibinfo{journal}{Rev. Mod. Phys.}\/},  {\it
  \bibinfo{volume}{93}\/}\bibinfo{issue}{(1)}, \bibinfo{pages}{015002}.

\bibitem[{Cowan et~al.(1991)Cowan, Thielemann \& Truran}]{cowan1991r}
\bibinfo{author}{Cowan, J.~J.}, \bibinfo{author}{Thielemann, F.~K.},  \&
  \bibinfo{author}{Truran, J.~W.} (\bibinfo{year}{1991}).
\newblock \bibinfo{title}{The $r$-process and nucleochronology}.
\newblock {\it \bibinfo{journal}{Phys. Rep.}\/},  {\it
  \bibinfo{volume}{208}\/}\bibinfo{issue}{(4-5)}, \bibinfo{pages}{267--394}.

\bibitem[{Dzhioev et~al.(2009)Dzhioev, Vdovin, Ponomarev \&
  Wambach}]{dzhioev2009thermal}
\bibinfo{author}{Dzhioev, A.~A.}, \bibinfo{author}{Vdovin, A.~I.},
  \bibinfo{author}{Ponomarev, V.~Y.} et~al. (\bibinfo{year}{2009}).
\newblock \bibinfo{title}{Thermal effects on electron capture for neutron-rich
  nuclei}.
\newblock {\it \bibinfo{journal}{Bull. Russ. Acad. Sci.: Phys.}\/},  {\it
  \bibinfo{volume}{73}\/}, \bibinfo{pages}{225--229}.

\bibitem[{Halbleib \& Sorensen(1967)}]{halbleib1967gamow}
\bibinfo{author}{Halbleib, S. J.~A.},  \& \bibinfo{author}{Sorensen, R.~A.}
  (\bibinfo{year}{1967}).
\newblock \bibinfo{title}{Gamow-teller beta decay in heavy spherical nuclei and
  the unlike particle-hole rpa}.
\newblock {\it \bibinfo{journal}{Nucl. Phys. A}\/},  {\it
  \bibinfo{volume}{98}\/}\bibinfo{issue}{(3)}, \bibinfo{pages}{542--568}.

\bibitem[{Hardy \& Towner(2009)}]{hardy2009superallowed}
\bibinfo{author}{Hardy, J.~C.},  \& \bibinfo{author}{Towner, I.~S.}
  (\bibinfo{year}{2009}).
\newblock \bibinfo{title}{Superallowed $0^+$→ $0^+$ nuclear $\beta$ decays: A
  new survey with precision tests of the conserved vector current hypothesis
  and the standard model}.
\newblock {\it \bibinfo{journal}{Phys. Rev. C. Nucl. Phys.}\/},  {\it
  \bibinfo{volume}{79}\/}\bibinfo{issue}{(5)}, \bibinfo{pages}{055502}.

\bibitem[{Heyde \& Wood(2011)}]{heyde2011publisher}
\bibinfo{author}{Heyde, K.},  \& \bibinfo{author}{Wood, J.~L.}
  (\bibinfo{year}{2011}).
\newblock \bibinfo{title}{Publisher’s note: Shape coexistence in atomic
  nuclei}.
\newblock {\it \bibinfo{journal}{Rev. Mod. Phys.}\/},  {\it
  \bibinfo{volume}{83}\/}\bibinfo{issue}{(4)}, \bibinfo{pages}{1655--1655}.

\bibitem[{Hirsch et~al.(1991)Hirsch, Staudt, Muto \&
  Klapdor}]{hirsch1991microscopic}
\bibinfo{author}{Hirsch, M.}, \bibinfo{author}{Staudt, A.},
  \bibinfo{author}{Muto, K.} et~al. (\bibinfo{year}{1991}).
\newblock \bibinfo{title}{Microscopic calculation of $\beta$+ ec decay
  half-lives with atomic numbers z= 10-30}.
\newblock {\it \bibinfo{journal}{Nucl. Phys. A.}\/},  {\it
  \bibinfo{volume}{535}\/}\bibinfo{issue}{(1)}, \bibinfo{pages}{62--76}.

\bibitem[{Hirsch et~al.(1993)Hirsch, Staudt, Muto \&
  Klapdor}]{hirsch1993microscopic}
\bibinfo{author}{Hirsch, M.}, \bibinfo{author}{Staudt, A.},
  \bibinfo{author}{Muto, K.} et~al. (\bibinfo{year}{1993}).
\newblock \bibinfo{title}{Microscopic predictions of $\beta$+/ec-decay
  half-lives}.
\newblock {\it \bibinfo{journal}{At. Data and Nucl. Data Tables}\/},  {\it
  \bibinfo{volume}{53}\/}\bibinfo{issue}{(2)}, \bibinfo{pages}{165--193}.

\bibitem[{Homma et~al.(1996)Homma, Bender, Hirsch, Muto \&
  Oda}]{homma1996systematic}
\bibinfo{author}{Homma, H.}, \bibinfo{author}{Bender, E.},
  \bibinfo{author}{Hirsch, M.} et~al. (\bibinfo{year}{1996}).
\newblock \bibinfo{title}{Systematic study of nuclear $\beta$ decay}.
\newblock {\it \bibinfo{journal}{Phys. Rev. C}\/},  {\it
  \bibinfo{volume}{54}\/}\bibinfo{issue}{(6)}, \bibinfo{pages}{2972}.

\bibitem[{Hosmer et~al.(2010)Hosmer, Schatz \& Aprahamian}]{hosmer2010half}
\bibinfo{author}{Hosmer, P.}, \bibinfo{author}{Schatz, H.},  \&
  \bibinfo{author}{Aprahamian, A. e.~a.} (\bibinfo{year}{2010}).
\newblock \bibinfo{title}{Half-lives and branchings for $\beta$-delayed neutron
  emission for neutron-rich co-cu isotopes in the $r$-process}.
\newblock {\it \bibinfo{journal}{Phys. Rev. C.—Nucl. Phys.}\/},  {\it
  \bibinfo{volume}{82}\/}\bibinfo{issue}{(2)}, \bibinfo{pages}{025806}.

\bibitem[{Ikeda(1963)}]{ikeda1963study}
\bibinfo{author}{Ikeda, M.} (\bibinfo{year}{1963}).
\newblock \bibinfo{title}{Study of interrelations between mechanisms at
  threshold}.
\newblock {\it \bibinfo{journal}{J. Opt. Soc. A.}\/},  {\it
  \bibinfo{volume}{53}\/}\bibinfo{issue}{(11)}, \bibinfo{pages}{1305--1313}.

\bibitem[{Klapdor et~al.(1984)Klapdor, Metzinger \& Oda}]{klapdor1984beta}
\bibinfo{author}{Klapdor, H.~V.}, \bibinfo{author}{Metzinger, J.},  \&
  \bibinfo{author}{Oda, T.} (\bibinfo{year}{1984}).
\newblock \bibinfo{title}{Beta-decay half-lives of neutron-rich nuclei}.
\newblock {\it \bibinfo{journal}{At. Data and Nucl. Data Tables}\/},  {\it
  \bibinfo{volume}{31}\/}\bibinfo{issue}{(1)}, \bibinfo{pages}{81--111}.

\bibitem[{Kondev et~al.(2021)Kondev, Wang, Huang, Naimi \&
  Audi}]{kondev2021nubase2020}
\bibinfo{author}{Kondev, F.~G.}, \bibinfo{author}{Wang, M.},
  \bibinfo{author}{Huang, W.~J.} et~al. (\bibinfo{year}{2021}).
\newblock \bibinfo{title}{The nubase2020 evaluation of nuclear physics
  properties}.
\newblock {\it \bibinfo{journal}{Chin. Phys. C.}\/},  {\it
  \bibinfo{volume}{45}\/}\bibinfo{issue}{(3)}, \bibinfo{pages}{030001}.

\bibitem[{Lalazissis et~al.(2005)Lalazissis, Nik{\v{s}}i{\'c}, Vretenar \&
  Ring}]{lalazissis2005new}
\bibinfo{author}{Lalazissis, G.~A.}, \bibinfo{author}{Nik{\v{s}}i{\'c}, T.},
  \bibinfo{author}{Vretenar, D.} et~al. (\bibinfo{year}{2005}).
\newblock \bibinfo{title}{New relativistic mean-field interaction with
  density-dependent meson-nucleon couplings}.
\newblock {\it \bibinfo{journal}{Phys. Rev. C.—Nucl. Phys.}\/},  {\it
  \bibinfo{volume}{71}\/}\bibinfo{issue}{(2)}, \bibinfo{pages}{024312}.

\bibitem[{Langanke \& Mart{\i}nez-Pinedo(2000)}]{langanke2000shell}
\bibinfo{author}{Langanke, K.},  \& \bibinfo{author}{Mart{\i}nez-Pinedo, G.}
  (\bibinfo{year}{2000}).
\newblock \bibinfo{title}{Shell-model calculations of stellar weak interaction
  rates: Ii. weak rates for nuclei in the mass range a= 45- 65 in supernovae
  environments}.
\newblock {\it \bibinfo{journal}{Nucl. Phys. A}\/},  {\it
  \bibinfo{volume}{673}\/}\bibinfo{issue}{(1-4)}, \bibinfo{pages}{481--508}.

\bibitem[{Lic{\u{a}} et~al.(2018)Lic{\u{a}}, Benzoni, Rodr{\'\i}guez \& et.
  al.}]{licua2018evolution}
\bibinfo{author}{Lic{\u{a}}, R.}, \bibinfo{author}{Benzoni, G.},
  \bibinfo{author}{Rodr{\'\i}guez, T.~R.} et~al. (\bibinfo{year}{2018}).
\newblock \bibinfo{title}{Evolution of deformation in neutron-rich ba isotopes
  up to a= 150}.
\newblock {\it \bibinfo{journal}{Phys. Rev. C.}\/},  {\it
  \bibinfo{volume}{97}\/}\bibinfo{issue}{(2)}, \bibinfo{pages}{024305}.

\bibitem[{Mei et~al.(2012)Mei, Xiang, Yao, Li \& Meng}]{mei2012rapid}
\bibinfo{author}{Mei, H.}, \bibinfo{author}{Xiang, J.}, \bibinfo{author}{Yao,
  J.~M.} et~al. (\bibinfo{year}{2012}).
\newblock \bibinfo{title}{Rapid structural change in low-lying states of
  neutron-rich sr and zr isotopes}.
\newblock {\it \bibinfo{journal}{Phys. Rev. C.—Nucl. Phys.}\/},  {\it
  \bibinfo{volume}{85}\/}\bibinfo{issue}{(3)}, \bibinfo{pages}{034321}.

\bibitem[{Meng et~al.(2006)Meng, Toki, Zhou, Zhang, Long \&
  Geng}]{meng2006relativistic}
\bibinfo{author}{Meng, J.}, \bibinfo{author}{Toki, H.}, \bibinfo{author}{Zhou,
  S.~G.} et~al. (\bibinfo{year}{2006}).
\newblock \bibinfo{title}{Relativistic continuum hartree bogoliubov theory for
  ground-state properties of exotic nuclei}.
\newblock {\it \bibinfo{journal}{Prog. Part. Nucl. Phys.}\/},  {\it
  \bibinfo{volume}{57}\/}\bibinfo{issue}{(2)}, \bibinfo{pages}{470--563}.

\bibitem[{Minato \& Hagino(2009)}]{minato2009beta}
\bibinfo{author}{Minato, F.},  \& \bibinfo{author}{Hagino, K.}
  (\bibinfo{year}{2009}).
\newblock \bibinfo{title}{$\beta$-decay half-lives at finite temperatures for
  n= 82 isotones}.
\newblock {\it \bibinfo{journal}{Phys. Rev. C.—Nucl. Phys.}\/},  {\it
  \bibinfo{volume}{80}\/}\bibinfo{issue}{(6)}, \bibinfo{pages}{065808}.

\bibitem[{M{\"o}ller et~al.(2016)M{\"o}ller, Sierk, Ichikawa \&
  Sagawa}]{moller2016nuclear}
\bibinfo{author}{M{\"o}ller, P.}, \bibinfo{author}{Sierk, A.~J.},
  \bibinfo{author}{Ichikawa, T.} et~al. (\bibinfo{year}{2016}).
\newblock \bibinfo{title}{Nuclear ground-state masses and deformations: Frdm
  (2012)}.
\newblock {\it \bibinfo{journal}{At. Data and Nucl. Data Tables}\/},  {\it
  \bibinfo{volume}{109}\/}, \bibinfo{pages}{1--204}.

\bibitem[{Moon et~al.(2017)Moon, Moon, Odahara \& et. al.}]{moon2017beta}
\bibinfo{author}{Moon, B.}, \bibinfo{author}{Moon, C.-B.},
  \bibinfo{author}{Odahara, A.} et~al. (\bibinfo{year}{2017}).
\newblock \bibinfo{title}{$\beta$-decay scheme of te 140 to i 140: Suppression
  of gamow-teller transitions between the neutron h 9/2 and proton h 11/2
  partner orbitals}.
\newblock {\it \bibinfo{journal}{Phys. Rev. C.}\/},  {\it
  \bibinfo{volume}{96}\/}\bibinfo{issue}{(1)}, \bibinfo{pages}{014325}.

\bibitem[{Muto et~al.(1989)Muto, Bender \& Klapdor}]{muto1989proton}
\bibinfo{author}{Muto, K.}, \bibinfo{author}{Bender, E.},  \&
  \bibinfo{author}{Klapdor, H.~V.} (\bibinfo{year}{1989}).
\newblock \bibinfo{title}{Proton-neutron quasiparticle rpa and charge-changing
  transitions}.
\newblock {\it \bibinfo{journal}{Z. Phys. A: Hadrons Nucl.}\/},  {\it
  \bibinfo{volume}{333}\/}, \bibinfo{pages}{125--129}.

\bibitem[{Muto et~al.(1992)Muto, Bender, Oda \& Klapdor}]{muto1992proton}
\bibinfo{author}{Muto, K.}, \bibinfo{author}{Bender, E.}, \bibinfo{author}{Oda,
  T.} et~al. (\bibinfo{year}{1992}).
\newblock \bibinfo{title}{Proton-neutron quasiparticle rpa with separable
  gamow-teller forces}.
\newblock {\it \bibinfo{journal}{Z. f{\"u}r Physik A. Hadr. Nucl.}\/},  {\it
  \bibinfo{volume}{341}\/}\bibinfo{issue}{(4)}, \bibinfo{pages}{407--415}.

\bibitem[{Nabi \& Klapdor(1999)}]{nabi1999weak}
\bibinfo{author}{Nabi, J.-U.},  \& \bibinfo{author}{Klapdor, H.~V.}
  (\bibinfo{year}{1999}).
\newblock \bibinfo{title}{Weak interaction rates of sd-shell nuclei in stellar
  environment calculated in the proton-neutron quasiparticle random phase
  approximation}.
\newblock {\it \bibinfo{journal}{At. Data and Nucl. Data Tables}\/},  {\it
  \bibinfo{volume}{71}\/}, \bibinfo{pages}{149}.

\bibitem[{Nabi \& Klapdor(2004)}]{nabi2004microscopic}
\bibinfo{author}{Nabi, J.-U.},  \& \bibinfo{author}{Klapdor, H.~V.}
  (\bibinfo{year}{2004}).
\newblock \bibinfo{title}{Microscopic calculations of stellar weak interaction
  rates and energy losses for fp-and fpg-shell nuclei}.
\newblock {\it \bibinfo{journal}{At. Data and Nucl. Data Tables}\/},  {\it
  \bibinfo{volume}{88}\/}\bibinfo{issue}{(2)}, \bibinfo{pages}{237--476}.

\bibitem[{Nabi et~al.(2022)Nabi, Ullah \& Khan}]{nabi2022investigation}
\bibinfo{author}{Nabi, J.~U.}, \bibinfo{author}{Ullah, A.},  \&
  \bibinfo{author}{Khan, Z.} (\bibinfo{year}{2022}).
\newblock \bibinfo{title}{Investigation of $\beta$-decay properties of
  neutron-rich cerium isotopes}.
\newblock {\it \bibinfo{journal}{Phys. Scr.}\/},  {\it
  \bibinfo{volume}{97}\/}\bibinfo{issue}{(12)}, \bibinfo{pages}{125302}.

\bibitem[{Nakamura(2010)}]{nakamura2010review}
\bibinfo{author}{Nakamura, K.} (\bibinfo{year}{2010}).
\newblock \bibinfo{title}{Review of particle physics}.
\newblock {\it \bibinfo{journal}{J. Phys. G: Nucl. Part. Phys.}\/},  {\it
  \bibinfo{volume}{37}\/}\bibinfo{issue}{(7 A)}.

\bibitem[{Nik{\v{s}}i{\'c} et~al.(2014)Nik{\v{s}}i{\'c}, Paar, Vretenar \&
  Ring}]{nikvsic2014dirhb}
\bibinfo{author}{Nik{\v{s}}i{\'c}, T.}, \bibinfo{author}{Paar, N.},
  \bibinfo{author}{Vretenar, D.} et~al. (\bibinfo{year}{2014}).
\newblock \bibinfo{title}{Dirhb—a relativistic self-consistent mean-field
  framework for atomic nuclei}.
\newblock {\it \bibinfo{journal}{Comput. Phys. Commun.}\/},  {\it
  \bibinfo{volume}{185}\/}\bibinfo{issue}{(6)}, \bibinfo{pages}{1808--1821}.

\bibitem[{Nik{\v{s}}i{\'c} et~al.(2002)Nik{\v{s}}i{\'c}, Vretenar, D.~Finelli
  \& Ring}]{nikvsic2002relativistic}
\bibinfo{author}{Nik{\v{s}}i{\'c}, T.}, \bibinfo{author}{Vretenar},
  \bibinfo{author}{D.~Finelli, P.} et~al. (\bibinfo{year}{2002}).
\newblock \bibinfo{title}{Relativistic hartree-bogoliubov model with
  density-dependent meson-nucleon couplings}.
\newblock {\it \bibinfo{journal}{Phys. Rev. C}\/},  {\it
  \bibinfo{volume}{66}\/}\bibinfo{issue}{(2)}, \bibinfo{pages}{024306}.

\bibitem[{Nik{\v{s}}i{\'c} et~al.(2008)Nik{\v{s}}i{\'c}, Vretenar \&
  Ring}]{nikvsic2008relativistic}
\bibinfo{author}{Nik{\v{s}}i{\'c}, T.}, \bibinfo{author}{Vretenar, D.},  \&
  \bibinfo{author}{Ring, P.} (\bibinfo{year}{2008}).
\newblock \bibinfo{title}{Relativistic nuclear energy density functionals:
  Adjusting parameters to binding energies}.
\newblock {\it \bibinfo{journal}{Phys. Rev. C. Nucl. Phys.}\/},  {\it
  \bibinfo{volume}{78}\/}\bibinfo{issue}{(3)}, \bibinfo{pages}{034318}.

\bibitem[{Nilsson(1955)}]{nilsson1955binding}
\bibinfo{author}{Nilsson, S.~G.} (\bibinfo{year}{1955}).
\newblock \bibinfo{title}{Binding states of individual nucleons in strongly
  deformed nuclei}.
\newblock {\it \bibinfo{journal}{Dan. Mat. Fys. Medd.}\/},  {\it
  \bibinfo{volume}{29}\/}\bibinfo{issue}{(CERN-55-30)}, \bibinfo{pages}{1--69}.

\bibitem[{Pfeiffer et~al.(2001)Pfeiffer, Kratz, Thielemann \&
  Walters}]{pfeiffer2001nuclear}
\bibinfo{author}{Pfeiffer, B.}, \bibinfo{author}{Kratz, K.~L.},
  \bibinfo{author}{Thielemann, F.~K.} et~al. (\bibinfo{year}{2001}).
\newblock \bibinfo{title}{Nuclear structure studies for the astrophysical
  $r$-process}.
\newblock {\it \bibinfo{journal}{Nuc. Phys. A.}\/},  {\it
  \bibinfo{volume}{693}\/}\bibinfo{issue}{(1-2)}, \bibinfo{pages}{282--324}.

\bibitem[{Ragnarsson \& Sheline(1984)}]{ragnarsson1984systematics}
\bibinfo{author}{Ragnarsson, I.},  \& \bibinfo{author}{Sheline, R.~K.}
  (\bibinfo{year}{1984}).
\newblock \bibinfo{title}{Systematics of nuclear deformations}.
\newblock {\it \bibinfo{journal}{Phys. Scrip.}\/},  {\it
  \bibinfo{volume}{29}\/}\bibinfo{issue}{(5)}, \bibinfo{pages}{385}.

\bibitem[{Reinhard(1989)}]{reinhard1989relativistic}
\bibinfo{author}{Reinhard, P.-G.} (\bibinfo{year}{1989}).
\newblock \bibinfo{title}{The relativistic mean-field description of nuclei and
  nuclear dynamics}.
\newblock {\it \bibinfo{journal}{Rep. Prog. Phys.}\/},  {\it
  \bibinfo{volume}{52}\/}\bibinfo{issue}{(4)}, \bibinfo{pages}{439}.

\bibitem[{Ring(1996)}]{ring1996relativistic}
\bibinfo{author}{Ring, P.} (\bibinfo{year}{1996}).
\newblock \bibinfo{title}{Relativistic mean field theory in finite nuclei}.
\newblock {\it \bibinfo{journal}{Prog. Part. Nucl. Phys.}\/},  {\it
  \bibinfo{volume}{37}\/}, \bibinfo{pages}{193--263}.

\bibitem[{Rodr{\'\i}guez-Guzm{\'a}n et~al.(2010)Rodr{\'\i}guez-Guzm{\'a}n,
  Sarriguren, Robledo \& Perez-Martin}]{rodriguez2010charge}
\bibinfo{author}{Rodr{\'\i}guez-Guzm{\'a}n, R.}, \bibinfo{author}{Sarriguren,
  P.}, \bibinfo{author}{Robledo, L.~M.} et~al. (\bibinfo{year}{2010}).
\newblock \bibinfo{title}{Charge radii and structural evolution in sr, zr, and
  mo isotopes}.
\newblock {\it \bibinfo{journal}{Phys. Lett. B.}\/},  {\it
  \bibinfo{volume}{691}\/}\bibinfo{issue}{(4)}, \bibinfo{pages}{202--207}.

\bibitem[{Shehzadi et~al.(2022)Shehzadi, Nabi \& Farooq}]{shehzadi2022half}
\bibinfo{author}{Shehzadi, R.}, \bibinfo{author}{Nabi, J.~U.},  \&
  \bibinfo{author}{Farooq, F.} (\bibinfo{year}{2022}).
\newblock \bibinfo{title}{Half-life prediction of some neutron-rich exotic
  nuclei prior to peak a=130}.
\newblock {\it \bibinfo{journal}{Phys. Scr.}\/},  {\it
  \bibinfo{volume}{97}\/}\bibinfo{issue}{(11)}, \bibinfo{pages}{115301}.

\bibitem[{Shergur et~al.(2005)Shergur, W{\"o}hr, Walters \& et.
  al.}]{shergur2005identification}
\bibinfo{author}{Shergur, J.}, \bibinfo{author}{W{\"o}hr, A.},
  \bibinfo{author}{Walters, W.~B.} et~al. (\bibinfo{year}{2005}).
\newblock \bibinfo{title}{Identification of shell-model states in sb 135
  populated via $\beta$-decay of sn 135}.
\newblock {\it \bibinfo{journal}{Phys. Rev. C. Nucl. Phys.}\/},  {\it
  \bibinfo{volume}{72}\/}\bibinfo{issue}{(2)}, \bibinfo{pages}{024305}.

\bibitem[{Si et~al.(2022)Si, Lozeva, Na{\"\i}dja \& et. al.}]{si2022new}
\bibinfo{author}{Si, M.}, \bibinfo{author}{Lozeva, R.},
  \bibinfo{author}{Na{\"\i}dja, H.} et~al. (\bibinfo{year}{2022}).
\newblock \bibinfo{title}{New $\beta$-decay spectroscopy of the te 137
  nucleus}.
\newblock {\it \bibinfo{journal}{Phys. Rev. C.}\/},  {\it
  \bibinfo{volume}{106}\/}\bibinfo{issue}{(1)}, \bibinfo{pages}{014302}.

\bibitem[{Staudt et~al.(1990)Staudt, Bender, Muto \&
  Klapdor}]{staudt1990second}
\bibinfo{author}{Staudt, A.}, \bibinfo{author}{Bender, E.},
  \bibinfo{author}{Muto, K.} et~al. (\bibinfo{year}{1990}).
\newblock \bibinfo{title}{Second-generation microscopic predictions of
  beta-decay half-lives of neutron-rich nuclei}.
\newblock {\it \bibinfo{journal}{At. Data and Nucl. Data Tables}\/},  {\it
  \bibinfo{volume}{44}\/}\bibinfo{issue}{(1)}, \bibinfo{pages}{79--132}.

\bibitem[{Tachibana et~al.(1990)Tachibana, Yamada \&
  Yoshida}]{tachibana1990improvement}
\bibinfo{author}{Tachibana, T.}, \bibinfo{author}{Yamada, M.},  \&
  \bibinfo{author}{Yoshida, Y.} (\bibinfo{year}{1990}).
\newblock \bibinfo{title}{Improvement of the gross theory of $\beta$-decay. ii:
  one-particle strength function}.
\newblock {\it \bibinfo{journal}{Prog. Theor. Phys.}\/},  {\it
  \bibinfo{volume}{84}\/}\bibinfo{issue}{(4)}, \bibinfo{pages}{641--657}.

\bibitem[{Takahara et~al.(1989)Takahara, Hino, Oda, Muto, Wolters, Glaudemans
  \& Sato}]{takahara1989microscopic}
\bibinfo{author}{Takahara, M.}, \bibinfo{author}{Hino, M.},
  \bibinfo{author}{Oda, T.} et~al. (\bibinfo{year}{1989}).
\newblock \bibinfo{title}{Microscopic calculation of the rates of electron
  captures which induce the collapse of o+ ne+ mg cores}.
\newblock {\it \bibinfo{journal}{Nucl. Phys. A.}\/},  {\it
  \bibinfo{volume}{504}\/}\bibinfo{issue}{(1)}, \bibinfo{pages}{167--192}.

\bibitem[{Takahashi \& Yamada(1969)}]{takahashi1969gross}
\bibinfo{author}{Takahashi, K.},  \& \bibinfo{author}{Yamada, M.}
  (\bibinfo{year}{1969}).
\newblock \bibinfo{title}{Gross theory of nuclear $\beta$-decay}.
\newblock {\it \bibinfo{journal}{Prog. Theor. Phys}\/},  {\it
  \bibinfo{volume}{41}\/}\bibinfo{issue}{(6)}, \bibinfo{pages}{1470--1503}.

\bibitem[{Typel(2005)}]{typel2005relativistic}
\bibinfo{author}{Typel, S.} (\bibinfo{year}{2005}).
\newblock \bibinfo{title}{Relativistic model for nuclear matter and atomic
  nuclei with momentum-dependent self-energies}.
\newblock {\it \bibinfo{journal}{Phys. Rev. C. Nucl. Phys.}\/},  {\it
  \bibinfo{volume}{71}\/}\bibinfo{issue}{(6)}, \bibinfo{pages}{064301}.

\bibitem[{Typel(2018)}]{typel2018relativistic}
\bibinfo{author}{Typel, S.} (\bibinfo{year}{2018}).
\newblock \bibinfo{title}{Relativistic mean-field models with different
  parametrizations of density-dependent couplings}.
\newblock {\it \bibinfo{journal}{Part.}\/},  {\it
  \bibinfo{volume}{1}\/}\bibinfo{issue}{(1)}, \bibinfo{pages}{3--22}.

\bibitem[{Walecka(1974)}]{walecka1974nuclear}
\bibinfo{author}{Walecka, J.~D.} (\bibinfo{year}{1974}).
\newblock \bibinfo{title}{Nuclear hydrodynamics in a relativistic mean field
  theory}.
\newblock {\it \bibinfo{journal}{Ann. Phys.}\/},  {\it \bibinfo{volume}{83}\/},
  \bibinfo{pages}{491}.

\bibitem[{Wood et~al.(1992)Wood, Heyde, Nazarewicz, Huyse \&
  Van~Duppen}]{wood1992coexistence}
\bibinfo{author}{Wood, J.~L.}, \bibinfo{author}{Heyde, K.},
  \bibinfo{author}{Nazarewicz, W.} et~al. (\bibinfo{year}{1992}).
\newblock \bibinfo{title}{Coexistence in even-mass nuclei}.
\newblock {\it \bibinfo{journal}{Phys. Rep.}\/},  {\it
  \bibinfo{volume}{215}\/}\bibinfo{issue}{(3-4)}, \bibinfo{pages}{101--201}.

\bibitem[{Wu et~al.(2020)Wu, Nishimura \& M{\"o}ller}]{wu2020beta}
\bibinfo{author}{Wu, J.}, \bibinfo{author}{Nishimura, S.},  \&
  \bibinfo{author}{M{\"o}ller, P. e.~a.} (\bibinfo{year}{2020}).
\newblock \bibinfo{title}{$\beta$-decay half-lives of 55 neutron-rich isotopes
  beyond the n= 82 shell gap}.
\newblock {\it \bibinfo{journal}{Phys. Rev. C}\/},  {\it
  \bibinfo{volume}{101}\/}\bibinfo{issue}{(4)}, \bibinfo{pages}{042801}.

\end{thebibliography}


\begin{thebibliography}{9}
\bibitem{Cow21} J. J. Cowan, C. Sneden, J. E. Lawler, A. Aprahamian, M. Wiescher, K. Langanke, G. Martínez-Pinedo, and F. K. Thielemann, . {\em Origin of the heaviest elements: The rapid neutron-capture process}.  {Rev. Mod. Phys.}, {\bf 93}, 015002 (2021).
\bibitem{Bur57} E. M. Burbidge,  G. R. Burbidge, and W. A. Fowler, {\em Synthesis of the elements in stars}. {Rev. Mod. Phys.},  {\bf 29}, 547 (1957).
\bibitem{Cow91} J. J. Cowan, F. K. Thielemann and  J. W. Truran,  {\em The r-process and nucleochronology}. {Phys. Rep.}, {\bf 208}, 267-394 (1991).
\bibitem{Pfe01} B. Pfeiffer, K. L. Kratz, F. K. Thielemann and W. B. Walters, {\em  Nuclear structure studies for the astrophysical r-process}. {Nucl. Phys. A}, {\bf 693}, 282-324 (2001).
\bibitem{Woo92} J. L. Wood, K. Heyde, W. Nazarewicz, M. Huyse and P. Van Duppen, . {\em Coexistence in even-mass nuclei}. {Phys. Rep.} {\bf 215}, 101-201 (1992).
\bibitem{Hey11} K. Heyde and J. L. Wood, . {\em Publisher?s Note: Shape coexistence in atomic nuclei.} {Rev. Mod. Phys.} {\bf 83}, 1467, (2011).
\bibitem{Xia12} J. Xiang, Z. P. Li, Z. X. Li, J. M. Yao and J. Meng, {\em Covariant description of shape evolution and shape coexistence in neutron-rich nuclei at N $\approx$ 60}. {Nucl. Phys. A}, {\bf 873}, 1-16 (2012).
\bibitem{Mei12} H. Mei, J. Xiang, J. M. Yao, Z. P. Li and J. Meng, {\em Rapid structural change in low-lying states of neutron-rich Sr and Zr isotopes.} {Phys. Rev. C-Nucl. Phys.}, {\bf 85}, 034321 (2012).
\bibitem{Ben09} M. Bender, K. Bennaceur, T. Duguet, P. H. Heenen, T. Lesinski and J. Meyer, {\em Tensor part of the Skyrme energy density functional. II. Deformation properties of magic and semi-magic nuclei.} {Phys. Rev. C-Nucl. Phys.}, {\bf 80}, 064302 (2009).
\bibitem{Rod11} R. Rodríguez-Guzmán, P. Sarriguren, L. M.  Robledo and S. Perez-Martin, {\em Charge radii and structural evolution in Sr, Zr, and Mo isotopes}. {Phys. Lett. B}, {\bf 691}, 202-207 (2010).
\bibitem{Lan00} K. Langanke and G. Mart?nez-Pinedo, {\em Shell-model calculations of stellar weak interaction rates: II. Weak rates for nuclei in the mass range A= 45-65 in supernovae environments}. {Nucl. Phys. A}, {\bf 673}, 481-508 (2000).
\bibitem{Bor96} I. N. Borzov, S. A. Fayans, E. Krömer and D. Zawischa,  {\em Ground state properties and $\beta$-decay half-lives near 132 Sn in a self-consistent theory}. {Z. Phys. A Had. Nucl.} {\bf 355}, 117-127 (1996).
\bibitem{Tak69} K. Takahashi and M. Yamada,  {\em Gross theory of nuclear $\beta$-decay}. {Prog. Theo. Phys.}, {\bf 41}, 1470-1503 (1969).
\bibitem{Mut89} K. Muto, E. Bender, and H. V. Klapdor,  { \em Proton-neutron quasiparticle RPA and charge-changing transitions}. {Z. Phys. A At. Nucl.}, {\bf 333}, 125-129 (1989).
\bibitem{Hos10} P. Hosmer, H. Schatz, A. Aprahamian, O. Arndt, R. R. C. Clement, A. Estrade, K. Farouqi, K. L. Kratz, S. N. Liddick, A. F. Lisetskiy and P. F. Mantica, {\em Half-lives and branchings for $\beta$-delayed neutron emission for neutron-rich Co?Cu isotopes in the r-process}. {Phys. Rev. C-Nucl. Phys.}, {\bf 82}, 025806 (2010).
\bibitem{Nab22} J.-U. Nabi, A. Ullah and Z. Khan, {\em Investigation of $\beta$-decay properties of neutron-rich Cerium isotopes}. {Phys. Scr.}, {\bf 97}, 125302 (2022).
\bibitem{She22} R. Shehzadi, J.-U. Nabi and F. Farooq,  {\em Half-life prediction of some neutron-rich exotic nuclei prior to peak A= 130}. {Phys. Scr.}, {\bf 97}, 115301 (2022).
\bibitem{Dzh08} A. A. Dzhioev, A. I. Vdovin, V. Y. Ponomarev and J. Wambach,  {\em Thermal effects on electron capture for neutron-rich nuclei}. {Bull. Russ. Acad. Sci.: Phys.}, {\bf 73}, 225-229 (2009).
\bibitem{Min09} F. Minato and K. Hagino,  {\em $\beta$-decay half-lives at finite temperatures for N = 82 isotones}. {Phys. Rev. C-Nucl. Phys.}, {\bf 80}, 065808 (2009).
\bibitem{Hal16} J. A. Halbleib Sr and R. A. Sorensen,  {\em Gamow-Teller beta decay in heavy spherical nuclei and the unlike particle-hole rpa}. {Nucl. Phys. A}, {\bf 98}, 542-568 (1967).
\bibitem{Kla84} H. V. Klapdor, J. Metzinger and T. Oda,  {\em Beta-decay half-lives of neutron-rich nuclei}. {At. Data and Nucl. Data Tab}, {\bf 31}, 81-111 (1984).
\bibitem{Sta90} A. Staudt, E. Bender, and K. Muto,  {\em Second-generation microscopic predictions of beta-decay half-lives of neutron-rich nuclei}. {At. Data and Nucl. Data Tab}, {\bf 44} 79-132 (1990).
\bibitem{Hir93} M. Hirsch, A. Staudt, and K. Muto,{\em Microscopic predictions of $\beta^+$/ec-decay half-lives}. {At. Data and Nucl. Data Tab}, {\bf 53}, 165-193 (1993).
\bibitem{Hom96} H. Homma, E.  Bender, E., and M. Hirsch, {\em Systematic study of nuclear $\beta$ decay}. {Phys. Rev. C}, {\bf 54}, 2972 (1996).
\bibitem{Nab99} J.-U. Nabi and   H. V. Klapdor-Kleingrothaus,  {\em Weak interaction rates of sd-shell nuclei in stellar environment calculated in the proton-neutron quasiparticle random phase approximation}. {At. Data and Nucl. Data Tab.}, {\bf 71}, 257 (1999).
\bibitem{Nab04} J.-U. Nabi and H. V.  Klapdor-Kleingrothaus,  {\em Microscopic calculations of stellar weak interaction rates and energy losses for fp-and fpg-shell nuclei}. {At. Data and Nucl. Data Tables}, {\bf 88(2)}, 237-476 (2004).
\bibitem{Tak89} M. Takahara, M.  Hino, T.  Oda, K. Muto, A. A. Wolters, P. W. M. Glaudemans and K. Sato,  {\em Microscopic calculation of the rates of electron captures which induce the collapse of O+ Ne+ Mg cores}. {Nucl. Phys. A}, {\bf 504}, 167-192 (1989).
\bibitem{Wu20} J. Wu, S. Nishimura, P. Möller,  {\em $\beta$-decay half-lives of 55 neutron-rich isotopes beyond the n= 82 shell gap}. {Phys. Rev. C}, {\bf 101}, 042801 (2020).
\bibitem{rei89} P. G. Reinhard, {\em The relativistic mean-field description of nuclei and nuclear dynamics}. {Rep. Prog. Phys.}, {\bf 52}, 439 (1989).
\bibitem{bay13}T. Bayram andA. H.  Yilmaz, {\em Table of ground state properties of nuclei in the RMF Model}. {Mod. Phys. Lett. A}, {\bf 28}, 1350068 (2013).
\bibitem{typ18} S. Typel,  {\em Relativistic mean-field models with different parametrizations of density dependent couplings}. {Part.}, {\bf 1} 3-22 (2018).
\bibitem{Mol16} P. Möller, A. J.  Sierk, T. Ichikawa, {\em Nuclear ground-state masses and deformations: Frdm (2012)}. {At. Data and Nucl. Data Tables}, {\bf 109}, 1-204 (2016).
\bibitem{wal74} J. D. Walecka,  {\em Nuclear hydrodynamics in a relativistic mean field theory}. {Ann. Phys.} {\bf 83}, 491, (1974).
\bibitem{rin96} P. Ring,  {\em Relativistic mean field theory in finite nuclei}. {Prog. Part. Nucl. Phys.} {\bf 37}, 193-263, (1996).
\bibitem{bog77} J. Boguta and A. R. Bodmer,  {\em Relativistic calculation of nuclear matter and the nuclear surface}. {Nucl. Phys. A}, {\bf 292}, 413-428, (1977).
\bibitem{men06} J. Meng, H. Toki, S. G.  Zhou, S. Q. Zhang, W. H. Long and  L. S. Geng,  {\em Relativistic continuum Hartree Bogoliubov theory for ground-state properties of exotic nuclei}. {Prog. Part. Nucl. Phys.}, {\bf 57}, 470-563, (2006).
\bibitem{nik02b}T. Nik?i?,, D. Vretenar, P. Finelli and P. Ring,  {\em Relativistic Hartree-Bogoliubov model with density-dependent meson-nucleon couplings}. {Phys. Rev. C}, {\bf 66}, 024306, (2002).
\bibitem{typ05}S. Typel,  {\em Relativistic model for nuclear matter and atomic nuclei with momentum-dependent self-energies}. {Phys. Rev. C-Nucl. Phys.}, {\bf 71}, 064301, (2005).

\bibitem{nik08}T. Nik?i?,, D. Vretenar and P.Ring,  {\em Relativistic nuclear energy density functionals: Adjusting parameters to binding energies}. {Phys. Rev. C-Nucl. Phys.}, {\bf 78}, 034318, (2008).

\bibitem{nik14}T. Nik?i?,, N. Paar, D.  Vretenar and P. Ring,  {\em DIRHB?A relativistic self-consistent mean-field framework for atomic nuclei}. {Comp. Phys. Comm.}, {\bf 185}, 1808-1821, (2014).
\bibitem{lal05} G. A. Lalazissis, T. Nik?i?, D. Vretenar and P. Ring,  {\em New relativistic mean-field interaction with density-dependent meson-nucleon couplings}. {Phys. Rev. C-Nucl. Phys.}, {\bf 71}, 024312 (2005).

\bibitem{Nil55} S. G. Nilsson,  {\em Binding states of individual nucleons in strongly deformed nuclei}. {Dan. Mat. Fys. Medd.}, 1-69 (1955).
\bibitem{Rag84} I. Ragnarsson and R. Sheline, {\em Systematics of nuclear deformations}. {Phys. Scr.}, {\bf 29}, 385 (1984).
\bibitem{Kon21} F. Kondev, M. Wang, W. Huang, et al.  {\em The nubase2020 evaluation of nuclear physics properties}. {Chin. Phys. C}, {\bf 45}, 030001, (2021).
\bibitem{Har09} J. C. Hardy, I. S. Towner,  {\em Super allowed $0^+\rightarrow 0^+$ nuclear $\beta$ decays: A new survey with precision tests of the conserved vector current hypothesis and the 236 standard model}. {Phys. Rev. C}, {\bf 79}, 055502, (2009).
\bibitem{Nak10} K. Nakamura,  {\em Review of particle physics}. {J. Phys. G: Nucl. Part. Phys.}, {\bf 37}(2010).
\bibitem{Mut92} K. Muto, E.  Bender, T.  Oda, et al.  {\em Proton-neutron quasiparticle rpa with separable gamow-teller forces}. {Z. Phys. A Had. Nucl.}, 
{\bf 341}, 407-415 (1992).
\bibitem{Hir91} M. Hirsch, A.  Staudt, K.  Muto, et al.  {\em Microscopic calculation of $\beta^+$ ec decay half-lives with atomic numbers z= 10?30}. {Nucl. Phys. A}, {\bf 535}, 62-76 (1991).
\bibitem{Gov71} N. Gove, M.  Martin,  {\em Log-f tables for beta decay}. {At. Data and Nucl. Data Tables}, {\bf 10}, 205-219, (1971).

\bibitem{Tac90} T. Tachibana, M. Yamada and Y. Yoshida, {\em Studies with one- and two-proton drip line nuclei.} {Prog. Theor. Phys.} \textbf{84}, 64, (1990).
\bibitem{She05} J. Shergur, A. Wöhr, W. B. Walters, K. L. Kratz, O. Arndt, B. A. Brown, J. Cederkall, I. Dillmann, L. M. Fraile, P. Hoff and A. Joinet, {\em Identification of shell-model states in $^{135}$Sb populated via $\beta^-$ decay of $^{135}$Sn.} {Phys. Rev. C} \textbf{72}, 024305, (2005).
\bibitem{Si22} M. Si, R. Lozeva, H. Naïdja, A. Blanc, J. M. Daugas, F. Didierjean, G. Duchêne, U. Köster, T. Kurtukian-Nieto, F. Le Blanc and P. Mutti, {\em New $\beta^-$ decay spectroscopy of the $^{137}$Te nucleus.}{Phys. Rev. C} \textbf{106}, 014302, (2022).
\bibitem{Moo17} B. Moon, A. Gargano, H. Naïdja, C. B. Moon, A. Odahara, R. Lozeva, S. Nishimura, C. Yuan, F. Browne, P. Doornenbal and G. Lorusso, {\em $\beta$-decay scheme of $^{140}$Te to $^{140}$I: Suppression of Gamow-Teller transitions between the neutron $h_9/2$ and proton $h_11/2$ partner orbitals,} {Phys. Rev. C} \textbf{96}, 014325, (2017).
\bibitem{Lic18} R. Lica, G. Benzoni, T. R. Rodríguez, M. J. G. Borge, L. M. Fraile, H. Mach, A. I. Morales, M. Madurga C. O. Sotty, V. Vedia and H. De Witte, {\em Evolution of deformation in neutron-rich Ba isotopes up to $A$ =150.} {Phys. Rev. C} \textbf{97}, 024305 (2018).

\end{thebibliography}
\end{document}